\documentstyle[11pt,aaspp4,epsf,flushrt]{article}

\slugcomment{\shortstack[r]{Submitted version, September 22, 1997 \\
HUPD-9717, KUNS-1462 }}

\lefthead{Yamamoto et al.}
\righthead{Evolution of Small Scale Cosmological Baryon 
Perturbations .....}

\begin{document}
\title{ Evolution of Small Scale Cosmological Baryon Perturbations\\
and\\
Matter Transfer Functions
}

\author{Kazuhiro Yamamoto\altaffilmark{1}}
\affil{Department of Physics, Hiroshima University,
Higashi-Hiroshima 739,~Japan}

\and

\author{Naoshi Sugiyama\altaffilmark{2} \& Humitaka Sato} 
\affil{Department of Physics, Kyoto University, Kyoto 606-01,~Japan}

\altaffiltext{1}{e-mail: yamamoto@astro.phys.sci.hiroshima-u.ac.jp}
\altaffiltext{2}{e-mail: naoshi@tap.scphys.kyoto-u.ac.jp}

\begin{abstract}

The evolution of small scale cosmological perturbations is carefully 
re-examined. Through the interaction with photons via electrons, 
baryon perturbations show interesting behavior in some physical 
scales. Characteristic features of the evolution of baryon density 
fluctuations are discussed. In CDM models, it is found a power-law 
growing phase of the small-scale baryon density fluctuations, which is 
characterized by the {\it terminal velocity}, after the
diffusion (Silk) damping and before the decoupling epoch.
Then, a transfer function for total matter density fluctuations
is studied by taking into account those physical processes.
An analytic transfer function is presented,
which is applicable for the entire range up to a solar mass scale
in the high$-z$ universe, 
and it is suitable also to the high baryon fraction models.

\end{abstract}

\keywords{cosmology: theory 
-- primordial density perturbations 
-- cosmic structure formation} 

\vfill
\thispagestyle{empty}
\newpage
\section{Introduction}
\def\II{{\rm I\hspace{-0.5mm}I}}
\def\III{{\rm I\hspace{-0.5mm}I\hspace{-0.5mm}I}}
\def\IV{{\rm I\hspace{-0.5mm}V}}
\def\VI{{\rm V\hspace{-0.5mm}I}}
\def\VII{{\rm V\hspace{-0.5mm}I\hspace{-0.5mm}I}}
\def\rmcomv{{}}
\def\N{N}
\def\aeq{{a_{\rm eq}}}
\def\keq{{k_{\rm eq}}}
\def\Th{\Theta_{2.726}}
\def\xe{{x_e}}
\def\ne{{n_{\rm e}}}
\def\fnu{{f_\nu}}
\def\The{\Theta_{2.726K}}
\def\yp{{y_{\rm p}}}
\def\nb{{n_{\rm b}}}
\def\MJ{{M_{\rm J}}}
\def\Msolar{{M_{\odot}}}
\def\Mpc{{\rm Mpc}}
\def\calA{{\cal A}}
\def\kphys{{k^{\rm phys}}}
\def\kcomv{{k^{\rmcomv}}}
\def\rhob{{\rho_{b}}}
\def\Omegab{{\Omega_{\rm b}}}
\def\Omegam{{\Omega_{\rm 0}}}
\def\Te{{T_{\rm b}}}
\def\cf{{c_{\rm f}}}
\def\cs{{c_{\rm s}}}
\def\ce{{c_{\rm e}}}
\def\me{{m_{\rm e}}}
\def\aova{{a\over a_0}}
\def\phys{{\rm phys}}
\def\deltab{{\delta_{\rm b}}}
\def\deltac{{\delta_{\rm c}}}
\def\deltam{{\delta_{\rm m}}}
The structure formation in the high-$z$ universe
is one of the most important problem in the fields of cosmology 
and astrophysics. The structure formation of
the Cold Dark Matter (CDM) dominated models
is well motivated from the recent 
cosmological observations, e.g., from 
the microwave background anisotropies
and the large scale structure of galaxies (\cite{WSS}; \cite{DGT}
; \cite{PD}),
although it may require some modifications, i.e., inclusion of 
cosmological constant, and the tilted initial power spectrum with 
the gravity wave mode (White, Scott, Silk, \& Davis 1995).
We expect that the power spectrum of density
fluctuations on scales $\gtrsim 1\rm Mpc$ 
will be measured precisely in near future, 
e.g., from the 2DF survey, the Slon Digital Sky Survey (SDSS) project 
(see, e.g., Strauss 1996; \cite{Loveday}) and the future 
cosmic microwave background experiments with satellites
\footnote{
 Planck home page
 {\tt http://astro.estec.esa.nl/SA-general/Projects/Cobras/cobras.html};
 MAP home page {\tt http://map.gsfc.nasa.gov}.
}
(see, e.g., \cite{ZSS}; \cite{JKKS}).
These observations will severely constrain the cosmological 
parameters and the initial density power spectrum.

On the other hand, the inflation scenario provides a successful 
mechanism to explain the origin of cosmological density
fluctuations.  It predicts almost scale invariant and Gaussian 
fluctuations (\cite{BST}).
In this scenario, we expect that
the initial density power spectrum extends up to 
very small scales.
The evolution of density perturbations 
due to the fluctuations on very small scales is 
especially important for the early formation of the bound objects 
such as population III stars or primordial sub-galaxies for
the hierarchical clustering scenario of the structure formation.
By studying this evolution up to the 
non-linear regime with taking into account
the heating and  cooling processes, 
we would be able to  learn the formation epoch and 
the formation process of the initial cosmic objects in the 
high--$z$ universe (e.g., \cite{OG}; \cite{GO}; Haiman \& Loeb 1997).

In this paper, we carefully re-examine the evolution of the small 
scale cosmological perturbations in the linear regime as 
an initial condition of the structure formation.
In cosmological models with the high baryon fraction, the evolution 
of baryon density fluctuations affects that of the total matter 
fluctuations. This derives an alternation of the standard matter 
transfer function of the total matter density fluctuations.
In particular, the power spectrum shows the damping on 
scales which are smaller than the Jeans scale 
at the decoupling epoch.  This damping is caused by 
the acoustic oscillation of baryon perturbations.
After describing these characteristic features of the evolution of  
baryon density fluctuations, we re-investigate the transfer
function for the total matter density fluctuations.

In \S~2, we give a brief review of the equations which describe 
the evolution of the cosmological perturbations. 
In \S~3, we describe various physical scales which are relevant to the 
characteristic features of the evolution of baryon density fluctuations.
The evolution of baryon density fluctuations
before and after the decoupling are discussed  in \S~3 and \S~4,
respectively.
The total matter transfer function is examined in \S~5.
In \S~6, the accuracy of the transfer function is 
shown by calculating statistical quantities such as $\sigma_8$.
A physical implication of our results is also demonstrated
by computing the formation rate of small bounded objects
in the high-$z$ universe based on the Press-Schechter theory.
\S~7 is devoted to summary and discussions.
In Appendix~A, we summarize the previous
results on the CDM perturbations in  small scales by 
Hu \& Sugiyama (1996) (hereafter HS96). 
In Appendix~B, we investigate the effect of the baryon 
thermal pressure on the transfer function on small scales
after the decoupling.

Throughout this paper, 
we work in units where $c=\hbar=k_B=1$.
And  we assume $T_0=2.726~{\rm K}$ as 
the cosmic microwave radiation temperature at present unless 
we explicitly include this valuable in equations.
\section{Review of Formalism}

Let us first summarize the equations which describe the
evolution of the cosmological density perturbations
on small scales.
We write the perturbed metric in the Newtonian gauge,
\begin{equation}
  ds^2=\biggl({a\over a_0}\biggr)^2 
  \Bigl(-(1+2\Psi)d\eta^2+(1+2\Phi)d{\bf x}^2\Bigr),
\end{equation}
where $\Psi$ is the perturbed gravitational potential,
$\Phi$ is the curvature perturbation,  
$a$ is the scale factor and the suffix $0$ indicates the present value.
As we are interested in the small scale cosmological perturbations,
we can assume the geometry of the universe to be flat.
We take into account the effect from 
the curvature term and the cosmological 
constant of the universe only near the present epoch through the change of 
the expansion rate.
Then, in the high redshift universe, 
the Friedmann equation is written as,
\begin{equation}
  H^2=\biggl({\dot a\over a }{a_0\over a}\biggr)^2=
  \biggl({a_0\over a}\biggr)^4 {\aeq+a\over \aeq+a_0} \Omega_0 H_0^2,
\end{equation}
where the dot denotes $\eta$-differentiation, 
$H_0$ is the Hubble constant, and $\aeq$ is the
scale factor at the matter-radiation equal-time.
The Friedmann equation can be integrated as 
$ \keq\eta/2\sqrt{2}=\bigl(\sqrt{1+a/\aeq}-1\bigr)$,
where we define
$  \keq^2=2\Omega_0H_0^2(a_0/\aeq)$, 
and 
\begin{eqnarray}
  && \keq=0.0948\times \sqrt{1-f_\nu} 
  \Omega_0 h^2 {\rm Mpc^{-1}},
\\
  &&{a_0\over \aeq}=4.04\times10^{4}(1-f_\nu)
  \Omega_0h^2,
\end{eqnarray}
with $f_\nu$ being the neutrino fraction in the energy density of 
radiations. For the massless standard neutrino model with three species,
$f_\nu\simeq0.405$. 

The evolution of 
photon and neutrino radiation fields is described by the the 
Boltzmann equation. The linearized evolution equation of the $k$-th Fourier 
mode of the photon temperature perturbations is (see e.g., \cite{KS}; Hu 
\& Sugiyama 1995, hereafter HS95),
\begin{equation}
  \dot \Theta +ik\mu(\Theta+\Psi)
  =-\dot\Phi+\dot\tau\Bigl(\Theta_0-\Theta
  -{1\over10}\Theta_2P_2(\mu)-i\mu V_{\rm b} \Bigr)~,
\end{equation}
where $\mu=\vec\gamma\cdot {\bf k} /\vert k\vert$, 
$\dot\tau=\ne\sigma_{\rm T}a/a_0$,
$\ne$ is the free electron number density, 
$\sigma_{\rm T}$ is the Thomson cross section,
$P_l(\mu)$ is the Legendre function,
$V_{\rm b}$ is the velocity of baryon perturbations,
$\Theta_l$ is defined by the multipole moment expansion of 
$\Theta(\eta,k,\vec\gamma)=\sum_{l=0}^\infty \Theta_l(\eta,k)~(-i)^l 
P_l(\mu)$, and $\vec\gamma$ is the directional vector of the photon 
momentum.
We can solve the equation using the multipole expansion method.
The perturbations for the neutrino radiation fields 
obey to a similar equation. 
Effects of 
neutrino perturbations, however, are not so important for the
structure formation 
on small scales. It should be taken into account only around 
the epoch of the horizon crossing (HS96).

In the fluid approximation,
the evolution of CDM density fluctuations is described as 
\begin{eqnarray}
  &&\dot\delta_c=-kV_c-3\dot\Phi,
\label{deltaceq}
\\
  &&\dot V_c=-{\dot a\over a}V_c+k\Psi,
\label{Vceq}
\end{eqnarray}
where $\delta_c$ and $V_c$ are the $k$-th Fourier mode of the 
perturbations of the CDM density contrast and the velocity, respectively.

In our gauge, the perturbed Einstein equation 
gives the extended Poisson equations,
\begin{equation}
  k^2\Phi=4\pi G\biggl({a\over a_0}\biggr)^2\rho_T
  \Bigl(\delta_T+3{\dot a\over a} (1+w_T)k^{-1}V_T\Bigr),
\end{equation}
and
\begin{equation}
  k^2(\Phi+\Psi)=-8\pi G\biggl({a\over a_0}\biggr)^2 p_T\Pi_T,
\end{equation}
where $\rho_T(1+w_T)V_T=\sum_x(\rho_x+p_x)V_x$,
$w_T=P_T/\rho_T$, $\rho_T\delta_T=\sum_x\rho_x\delta_x$, 
the stress anisotropy is $\Pi_T={12}(\Theta_2+N_2)/5$, $\Theta_2$ and 
$N_2$ are the quadrupole anisotropies of the photon and the neutrino 
temperature perturbations.

We next consider the baryon-electron system.
Electron({\rm e}), neutral and ionized hydrogen({\rm H}) 
and helium({\rm He}) are the particle species of the
baryon-electron system.
The number densities for each species are written as 
\begin{equation}
  n_{\rm e}=\xe\Bigl(1-{\yp\over2}\Bigr)\nb,
\hspace{7mm}
  n_{\rm H}=\Bigl(1-{\yp}\Bigr)\nb,
\hspace{7mm}
  n_{\rm He}={\yp\over4}\nb,
\end{equation}
where $\yp$ is the primordial helium mass fraction,
$n_{\rm H}$ and $n_{\rm He}$ are the number densities
of hydrogen and helium, respectively,
$\nb=n_{\rm H}+4n_{\rm He}$ is the total baryon number density,
and $\xe=\ne/(n_{\rm H}+n_{\rm He})$ is the electron ionization fraction.
We take a single fluid approximation for this baryon and electron system.
Then the evolution equations of perturbations of the
baryon-electron fluid can be written as 
\begin{eqnarray}
  &&\dot\delta_b=-kV_b-3\dot\Phi,
\label{dotdeltab}
\\
  &&\dot V_b=-{\dot a\over a}V_b+k\cs^2\delta_b+k\Psi
  +\dot\tau(\Theta_1-V_b)/R,
\label{Vbeq}
\end{eqnarray}
where $\cs$ is the sound velocity of the baryon-electron fluid,
and 
$R \equiv {3\rho_b/ 4\rho_\gamma}={3\Omegab/4\Omega_0}(1-\fnu)^{-1}(a/\aeq)$.

In the previous paper (Yamamoto, Sugiyama, \& Sato 1997, hereafter YSS), 
we gave useful formulation for the small scale baryon perturbations. 
In particular we investigated how the Compton interaction between the
electrons and the background photons determines the sound 
velocity of the baryon perturbations after the recombination.
According to the previous result,
the effective sound velocity of the baryon-electron fluid
after the recombination can be determined by solving the
dispersion relation, 
\begin{equation}
  -i\eta_{\rm E}\omega(\omega^2-\cf^2 k^2)+ \omega^2-\ce^2 k^2=0 .
\label{dispersionA}
\end{equation}
Here the real part of $\omega/k$ can be regarded as the effective 
sound velocity, i.e., $\cs={\rm Re} \bigl[{\omega/ k}\bigr]$,
$\eta_{\rm E}$ is the energy transfer time of 
the Compton interaction, 
\begin{equation}
  \eta_{\rm E}^{-1}={8\over3}{\aova}{\xe\Bigl(1-\yp/2\Bigr)
  \sigma_{\rm T}\rho_{\gamma}\over\me
  \Bigl(1+\xe-(\xe+3/2)\yp/2\Bigr)}~,
\end{equation}
$\cf^2={5}{P_0/3\rhob}$, $\ce^2={P_0/\rhob}$, $P_0$ and $\rhob$ 
are the pressure and the energy density of the baryon-electron 
fluid, respectively.
Note that $\cf$ is the sound speed for an adiabatic process 
and $\ce$ for an isothermal process.
The effective adiabatic index $\gamma$ can be defined as 
$\gamma=\cs^2/\ce^2$. As is clear from equation (\ref{dispersionA}),
the effective sound velocity $\cs$ (adiabatic index $\gamma$) is ruled 
by the ratio of the Compton energy transfer time scale $\eta_{\rm E}$ 
to the sound oscillation time scale.
Namely, if $\eta_E \omega \ll 1$, then $\omega \simeq c_e k$ and
$\gamma \simeq 1$.  
On the other hand, if 
$\eta_E \omega \gg 1$, $\omega \simeq c_{\rm f} k$ and
$\gamma \simeq 5/3$. 
Therefore the sound velocity depends on the wavelength of the perturbations,
and $\gamma$ is changed from $1$ to $5/3$ as the universe expands.

\section{Evolution of the baryon perturbations 
         before the decoupling}

In this section, we describe the evolution of the 
cosmological baryon perturbations  on  small scales before 
the decoupling.   The decoupling epoch (or the drag epoch) 
$\eta_d$ of baryon perturbations is 
defined by $\tau_d(\eta_d,\eta_0) =1 $, where 
\begin{equation}
\tau_d(\eta_1,\eta_2) \equiv \int_{\eta_2}^{\eta_2} d\eta 
{\dot\tau(\eta) \big/ R(\eta)} ~.
\end{equation}
One can find the analytic fitting formula of this epoch in HS96.  
Note that this epoch is not necessarily equal to 
the last scattering epoch of photons
which is defined by $\tau(\eta_{rec}, \eta_0)=1$.

   From the Compton interaction between electrons and photons,
there are interesting  varieties of 
physical scales for the baryon-electron fluid.
In the previous paper (\cite{YSS}),
we summarized the details of the physical scales 
which are relevant to the evolution of baryon fluctuations.
In this paper we focus on the small scale density perturbations,
and describe the evolution using the physical scales. 
We define $k^{\rm phys}$  ($\lambda^{\rm phys} \equiv 2\pi/k^{\rm phys}$) 
as a physical 
wave number 
(wavelength),  and  $k$ ($\lambda$) as a comoving wave number
(wavelength). They are related as $k=(a/a_0) k^{\rm phys}$ 
~$\bigl(~\lambda=(a_0/a)\lambda^{\rm phys}~\bigr)$.

\subsection{Horizon crossing and Jeans oscillation }

The first characteristic scale is the horizon crossing scale in the 
evolution of perturbations. 
We define the horizon crossing scale 
by $1/k_{\rm H}=\eta$, and the corresponding baryon mass scale by 
$M_{\rm H}=4\pi\rhob(\pi/k_{\rm H}^{\phys})^3/3$. Then we have 
$M_{\rm H}=7.9\times 10^{30}(1+z)^{-3}(1-\fnu)^{3/2}\Omegab h^2~\Msolar$,
in the radiation dominated stage.
According to this definition, the horizon crossing occurs at the redshift 
\begin{equation}
1+z_{\rm H}=2.0\times10^{10} (1-\fnu)^{1/2} (\Omegab h^2)^{1/3}
  \bigl({M/ \Msolar}\bigr)^{-1/3} ~ ,
\end{equation}
for  perturbations with the baryon mass scale $M$.
Since we have assumed that the perturbations cross the horizon in 
the radiation dominated stage, this expression 
is applicable for 
$M\ll 1.2\times10^{17} (1-\fnu)^{-3/2} \Omegab h^2(\Omega_0 h^2)^{-3}
~\Msolar$.

The next important scale is the Jeans scale. Since the coupling
between baryons and photons is tight due to the Compton interaction 
sufficiently before recombination, 
the photon-pressure prevents perturbations from collapsing.
Therefore the baryon-photon perturbations oscillate as an acoustic wave
inside the Jeans scale.
We define the Jeans wavelength (wave number) before recombination by
$\lambda_{\rm J}^{\rm phys}={2\pi/k_{\rm J}^{\rm phys}} \equiv
  ({\pi \cs^2/ G(\rho_{\rm m}+\rho_\gamma)})^{1/2}$,
where $\rho_{\rm m}=\rhob+\rho_{\rm c}$ and $\rho_{\rm c}$ is 
the energy density of CDM. Defining the Jeans mass as 
$M_{\rm J}={4\pi\rhob}({\lambda_{\rm J}^{\rm phys}/2})^3/3=
{4\pi\rhob}({\pi/k_{\rm J}^{\rm phys}})^3/3$, we get
\begin{equation}
  M_{\rm J}=8.6\times10^{29} (1+z)^{-3} \Omegab h^2 ~\Msolar,
\end{equation}
in the radiation dominated stage.
Then the Jeans crossing occurs at redshift,
\begin{equation}
1+z_{\rm J}=9.4\times10^{9}  (\Omegab h^2)^{1/3}
\bigl({M/ \Msolar}\bigr)^{-1/3} ~ ,
\label{jeansin}
\end{equation}
for the perturbations with the baryon mass $M$, which is right after the
horizon crossing epoch.
Since we have assumed perturbations cross the Jeans scale in 
the radiation dominated stage, this expression 
is applicable for 
$M\ll  10^{16} \Omegab h^2(\Omega_0 h^2)^{-3}~\Msolar$.
When the coupling between baryons and photons is tight,
the baryon-photon system behaves as a single fluid.
The oscillation is expressed in the analytic form, 
\begin{equation}
  \deltab={9\over2}\biggl(1+{2\over5}\fnu\biggr)^{-1}\Phi(0,k)
  (1+R)^{-1/4}\cos(k r_{\rm s}),
\label{acoustic}
\end{equation}
for adiabatic perturbations (HS95;HS96),
where $r_{\rm s}=\int_0^{\eta} {d\eta'}/\sqrt{3(1+R)}$, 
and $\Phi(0,k)$ is  the initial curvature perturbation.

\begin{figure}[t]
\centerline{\epsfxsize=10cm \epsffile{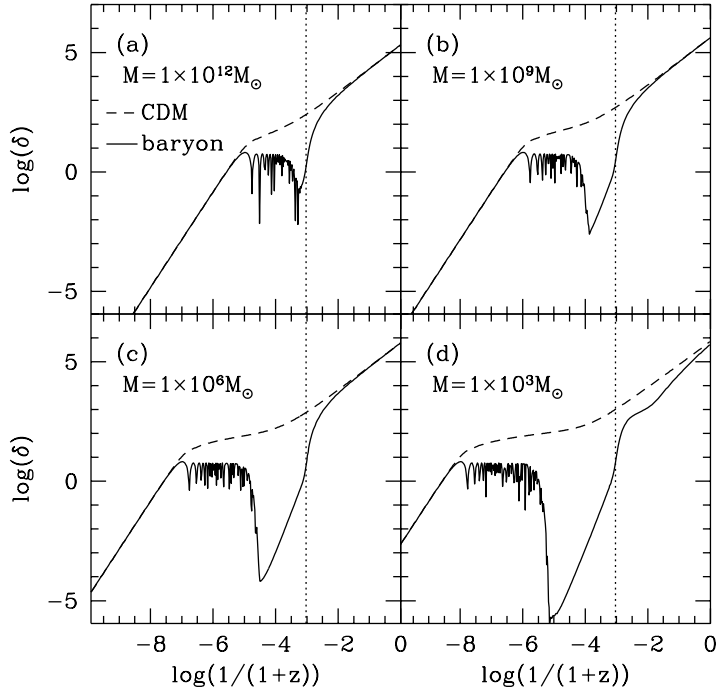}}
\caption{\protect\footnotesize
  Typical numerical evolution of baryon and the CDM density
  fluctuations, $\deltab$ (solid lines) and $\deltac$ (dashed lines), 
  with baryon mass scales 
  $10^{12}\Msolar, ~10^{9}\Msolar, ~10^6\Msolar$, and $10^3\Msolar$, 
  respectively. These mass scales correspond to 
  the wave numbers $k=9.65 \times 10^{-1}, \times 10^{0}, \times 10^{1},$
  and $\times 10^{2}$Mpc$^{-1}$, respectively.
  Note that here we employ the gauge to set the total matter
  center-of-mass frame, 
  instead of the Newtonian gauge for the convenience of
  numerical calculations.  The behavior of fluctuations 
  inside the horizon does not 
  depend on the gauge choice.
  We chose the cosmological parameters,  
  $h=0.5$, $\Omega_0=1.0$, and $\Omegab=0.1$.
  The dotted line shows the decoupling time of baryons and photons.
  We see the power-law growing phase of baryon density fluctuations
  after the diffusion damping before the decoupling. 
  After the decoupling the baryon fluctuations catch up with 
  CDM fluctuations through the gravitational free fall, 
  unless the perturbations are smaller than the Jeans scale (see panel (d)).
 }
\label{Fig.1}
\end{figure}
   Figure~\ref{Fig.1} shows the numerical  evolution of baryon and 
CDM density fluctuations with various 
baryon mass scales, $10^{12}\Msolar, ~10^{9}\Msolar, ~10^6\Msolar,$ and  
$10^3\Msolar$.
It is apparent from the figure that baryon density fluctuations
start oscillating after the Jeans crossing.
It is also well known that CDM density fluctuations grow logarithmically 
after the horizon crossing in the radiation dominant regime (see e.g., HS96).
Moreover, this figure shows some prominent features of the evolution of 
baryon density fluctuations, i.e., damping, re-growth and catching up with 
CDM density fluctuations.  
These are discussed in the following subsections and 
\S 4.

\subsection{Diffusion damping and the breaking scale 
of the tight coupling approximation}

It is well known that the photon diffusion process erases the 
baryon-photon perturbations on small scales 
(Silk 1968;~Sato 1971;~Weinberg 1971).
This damping length is the random walk distance
of a photon scattered by electrons through the Compton interaction. 
The mean free path of a photon is $1/\dot\tau$.
The integrated distance that the photon proceeds 
is the horizon scale $\eta$ in terms of the comoving length. 
Hence this damping scale is roughly expressed as 
$l\simeq\sqrt{\eta/\dot\tau}$. In a more detailed treatment,
we define the
damping scale by $k_{\rm D}$, where ${k^{-2}_{\rm D}}
=\int_0^\eta d\eta' {(R^2+8(1+R)/9)/(6\dot\tau (1+R)^2)}$  (see e.g., HS95).
It is known that the exponential damping factor, 
$\exp[-{k^2/ k^2_{\rm D}}]$ is multiplied to the right hand 
side of equation (\ref{acoustic}) to describe this damping feature.
Correspondingly, the baryon mass scale of the diffusion damping is
defined by $M_{\rm D}={4\pi\rhob}({\pi/k_{\rm D}^{\rm phys}})^3/3$, 
which reduces to
\begin{equation}
  M_{\rm D}={1.4\times10^{27} (1+z)^{-9/2}}
  (1-\yp/2)^{-3/2}(1-\fnu)^{3/4} 
  (\Omegab h^2)^{-1/2}  ~{\Msolar}.
\end{equation}
   For the perturbations with the baryon mass scale $M$,
this damping occurs at redshift
\begin{equation}
1+z_{\rm D}=1.1\times10^{6}(1-f_\nu)^{1/6}(1-y_{\rm p})^{-1/3}
(\Omegab h^2)^{-1/9} ({M/ \Msolar})^{-2/9} ~ . 
\end{equation}
Here we have assumed this regime to be 
sufficiently before recombination.

Now let us examine the validity of the tight coupling approximation
for the baryon-photon system.  When the photon mean free path 
crosses over the wavelength of the perturbations, the tight coupling 
approximation for the baryon-photon system cannot be applicable anymore. 
We define this breaking scale of the tight coupling approximation by 
$1/k_{\rm BR}^{\rm phys}=\lambda^{\rm phys}_{\rm BR}/2\pi
=1/n_e\sigma_T$. 
The corresponding baryon mass scale is defined by 
$M_{\rm BR}={4\pi} \rhob(\pi/k_{\rm BR}^{\rm phys})^3/3$, which reduces to
\begin{equation}
  M_{\rm BR}=2.9\times 10^{27} (1+z)^{-6} (1-\yp/2)^{-2} 
  (\Omegab h^2)^{-2}~{\Msolar},
\label{breakingmass}
\end{equation}
where we have used $\xe=1$ assuming the regime sufficiently before
recombination. Accordingly, the tight coupling approximation breaks down at
redshift 
\begin{equation}
1+z_{\rm BR}=3.8\times10^{4}(1-y_{\rm p})^{-1/2}(\Omegab h^2)^{-1/3}
({M/ \Msolar})^{-1/6}~ , 
\end{equation}
for the perturbations with the baryon mass scale $M$. 
Extrapolating equation (\ref{breakingmass}) until the decoupling epoch
($z \simeq 1000$),
we obtain
$M_{\rm BR}=2.9\times 10^{9} \left((1+z)/1000\right)^{-6}
(1-\yp/2)^{-2}(\Omegab h^2)^{-2}~{\Msolar}$.

\begin{figure}[t]
\centerline{\epsfxsize=10cm \epsffile{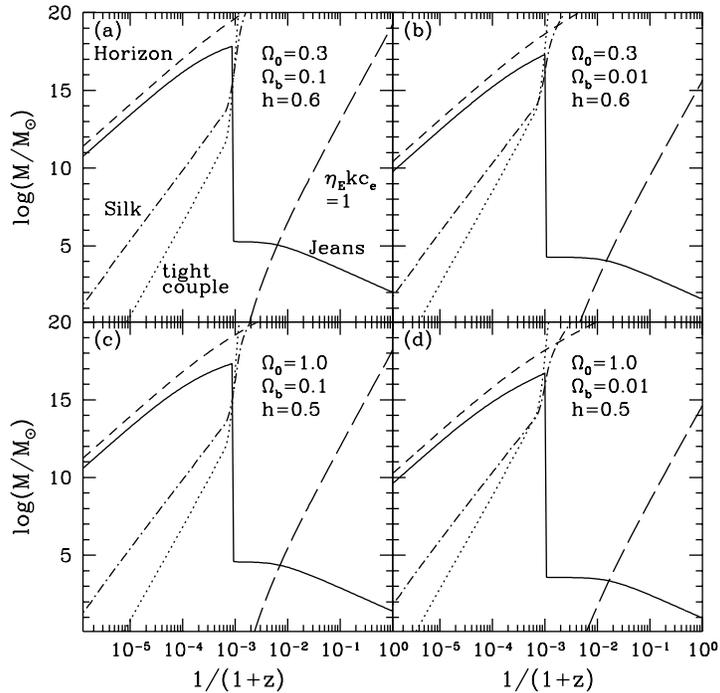}}
\caption{\protect\footnotesize
 Physical mass scales of baryon perturbations for several different 
 cosmological parameters. 
 In this figure, `Horizon' (short dashed lines), 'Jeans' (solid lines),
 'Silk' (dot-dashed lines) and  'tight couple' (dotted lines) represent
 the horizon scale, the Jeans scale, the diffusion (Silk) damping scale, and
 the breaking scale of the tight coupling approximation of 
 baryon and photon fluids.  Long dashed lines are the lines on which 
 the Compton energy time is equal to the sound oscillation time scale,
 i.e., $\eta_{\rm E} k c_{\rm e}= 1$. 
}
\label{Fig.2}
\end{figure}
In Figure~\ref{Fig.2} we show the physical baryon mass scales 
described in the above.
As can be read from this figure, the tight coupling 
approximation breaks down after the diffusion damping but 
sufficiently before recombination on small scales. 
Next we study
the effect of this tight coupling breaking 
on the evolution of density fluctuations. 

\subsection{Evolution of baryon fluctuations by the terminal velocity}
\begin{figure}[t]
\centerline{\epsfxsize=7cm \epsffile{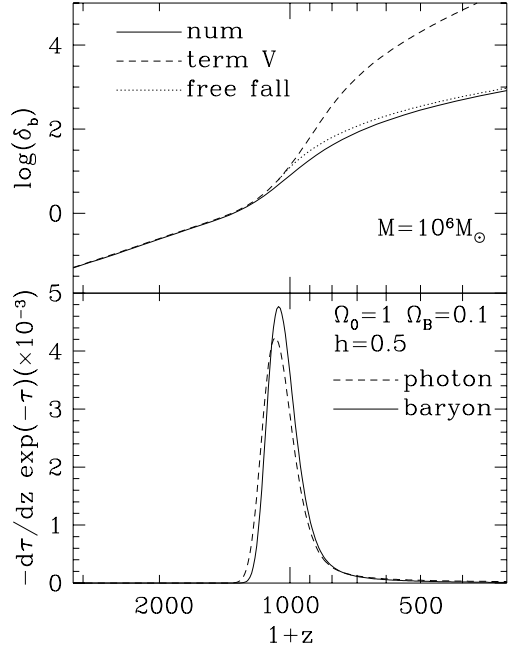}}
\caption{\protect\footnotesize
  Evolution of baryon density fluctuations and the visibility
  function around the decoupling epoch.
  The upper panel shows the evolution of baryon density fluctuations
  with the baryon mass scale  $~10^6\Msolar$ which corresponds to   
  $k=96.5\rm Mpc^{-1}$ (see panel (c) of Figure 1).  
  The cosmological parameters are same as 
  Figure~1.
  In this panel, `num' (a solid line), 'term V' (a dashed line), 
  and 'free fall' (a dotted line) represent
  fully numerical computation,
  the semi-analytic computation with setting $V_b=k\Psi R/\dot\tau$,
  i.e., fluctuations induced by the terminal velocity, 
  and the semi-analytic computation with setting $V_b$ to evolve 
  via gravitational free fall
  after the decoupling.
  The lower panel shows the corresponding 
  visibility functions. 
  Note that the last scattering epoch, which is the location of the peak 
  of the visibility function, 
  of photons (a dashed line) and 
  baryons (a solid line) are not completely the same. We refer the latter
  as a decoupling epoch in this paper.  
  It is shown in this figure 
  that  the semi-analytic calculation starts to deviate
  from the numerical calculations at which the decoupling
  process takes place (see the thickness of the visibility function).  
}
\label{Fig.3}
\end{figure}

In Figure~\ref{Fig.1}, which is obtained by numerical computations,
we find the power-law growing phase of baryon fluctuations 
after the diffusion damping but sufficiently before the decoupling 
on small scales.
This evolution of baryon fluctuations is characterized as follows.
Equation (\ref{Vbeq}) has the formal solution
\begin{equation}
  aV_b(\eta)=\int_0^\eta d\eta' a\Bigl( {\dot\tau\over R} \Theta_1
  +k c_b^2 \delta_b + k \Psi  \Bigr) e^{-\tau_d(\eta',\eta)},
\label{formalsol}
\end{equation}
where $\tau_d(\eta',\eta)= \int_{\eta'}^{\eta} d\tilde\eta 
{\dot\tau(\tilde\eta) \big/ R(\tilde\eta)}$.
Since the diffusion damping erases baryon-photon perturbations,
the dominant term in the integrant of the right hand side 
of equation (\ref{formalsol}) 
is the gravitational force term. Therefore we can approximate 
$ aV_b(\eta)\simeq\int_0^\eta d\eta' a    k \Psi  e^{-\tau_d(\eta',\eta)}$.
Before the decoupling, the optical depth 
$\tau_d(\eta',\eta)$ is a quite large number, and the time variation 
of $e^{-\tau_d(\eta',\eta)}$ is extreamly rapid. 
Thus we approximate as $e^{-\tau_d(\eta',\eta)}\simeq 
e^{-(\eta-\eta')\dot\tau(\eta)/R(\eta)}$, 
and neglect the time dependence
of $\Psi$, which derives
\begin{equation}
  V_b(\eta)\simeq k \Psi {R\over \dot\tau}~.
\label{terminalv}
\end{equation}
We can verify that this relation holds quite well 
during the power-law growing phase as is shown in Figure~\ref{Fig.3}.
We will discuss this figure in detail later in this subsection.
The above relation is also obtained by equating 
the gravitational force due to the potential of  CDM 
fluctuations, $k\Psi$, 
and the friction force due to the interaction with  
background photons, $\dot\tau V_{\rm b}/R$, in the right hand side of 
equation (\ref{Vbeq}).
Thus this relation implies the balance of these two forces 
for baryon perturbations. Therefore
this baryon velocity can be referred as the {\it terminal velocity}.
This is the result of the breaking of the tight coupling approximation.
After the epoch $\tau_d\lesssim1$, the baryon velocity is induced 
by the gravitational free fall, i.e., 
$V_b(\eta)=a^{-1}\left(
\int_{\eta_{\rm d}}^\eta d\eta'ak\Psi + a_{\rm d}V_b(\eta_{\rm d})\right) $.
Therefore the power-law growth by the terminal velocity 
lasts by the end of the baryon drag epoch.

Let us discuss this growing feature of the evolution by the terminal velocity 
more quantitatively.
In the matter dominant stage, the time variation of 
the curvature perturbation can be neglected.
In this case, baryon density fluctuations are given by
$ \deltab(\eta)
\simeq -k^2 \Psi(\eta)\int^\eta d\eta' {R/ \dot\tau}$, 
from equations (\ref{dotdeltab}) and (\ref{terminalv}).
Assuming $x_e=1$, we have $R/\dot\tau=3.0\times 
10^4 \Omegab h^2/(n_e\sigma_T)$. Then we can derive
\begin{equation}
  \deltab(\eta)\simeq-0.32(1-f_\nu)(1-\yp/2)^{-1}(\Omega_0 h^2)^{3/2} 
  \biggl({1000\over 1+z}\biggr)^{7/2} 
  \biggl({k\over \keq}\biggr)^2 \Psi(\eta) .
\label{terminaldb}
\end{equation}
It should be noticed that this relation is applicable between 
the breaking epoch of the tight coupling and the decoupling epoch.
The ratio of baryon density fluctuations to CDM ones is written as
\begin{equation}
  {\deltab(\eta) \over\deltac(\eta)} \simeq
  0.7\times 10^{-2} {(\Omega_0-\Omegab)\Omega_0^{-1/2}h}
  \biggl({1000\over 1+z}\biggr)^{5/2}~,
\label{ratio}
\end{equation}
where we took $\yp=0.23$,  since CDM density fluctuations
can be written as 
\begin{equation}
  \deltac(\eta)\simeq -{5.4\times 10}
  (1-\fnu)
  {\Omega_0^2h^2\over \Omega_0-\Omegab} \biggl({1000\over 1+z}\biggr) 
  \biggl({k\over \keq}\biggr)^2 \Psi(\eta) ~ , 
\end{equation}
which is obtained from the relation 
$k^2\Psi\simeq -4\pi G (a/a_0)^2\rho_{\rm c} \deltac$.
   From equation (\ref{ratio}), it is found that 
the ratio of baryon density fluctuations to 
CDM ones does not depend on the wave number $k$ before the decoupling epoch.

Since the above estimation is rather crude, we 
investigate the behavior around the decoupling epoch in more detail. 
Equation (\ref{terminaldb}) is a good approximation 
for $\deltab$ before the decoupling epoch.
However, we cannot employ equation (\ref{terminaldb}) 
during the decoupling epoch, 
because the decoupling process occurs in 
the finite time duration and 
our assumption of 
the fully ionization 
breaks down. Figure~\ref{Fig.3} shows the evolution of baryon 
fluctuations and the corresponding baryon and photon visibility functions
(HS96).
In the upper panel, the solid line shows the evolution obtained from 
the fully numerical computation. The dashed line shows the 
evolution of the baryon fluctuations with setting the baryon 
velocity to be the terminal velocity, i.e., $V_b=k\Psi R/\dot\tau$. 
The dotted line shows the evolution with setting that 
$V_b=k\Psi R/\dot\tau$ until the decoupling $z_d$ defined in HS96,
and setting $V_b$ to follow the free fall solution, i.e., 
$\dot V_b=-V_b(\dot a/a) + k\Psi$
after the decoupling.
   From this figure, it is apparent that our simple picture such that 
the small scale baryon fluctuations grow
via the terminal velocity until the decoupling epoch, 
and they grows via gravitational free fall after that, 
works remarkably well before the decoupling epoch and 
works fairly well after the decoupling epoch.  
However, there appears a small discrepancy  
during the decoupling epoch, i.e., within the thickness of 
the visibility function (see the lower panel).
The evolution of baryon fluctuations accelerates 
as the decoupling process begins.
However the actual acceleration is not as fast as the  
one which is given by the terminal velocity  
(see the dashed line).
This is because  the friction between baryons and photons is getting smaller
as the universe becomes transparent.  If the 
friction force and gravitational force were still balanced, 
unrealistically large velocity would be required.
Therefore we overestimate the velocity if we employ the terminal
velocity during and after the decoupling epoch.  
This is why the 
dashed 
line exceeds the numerical one (the solid line) at 
the decoupling time $z_d$ which is the peak location of 
the baryon visibility function.  Moreover, we have entirely ignored the 
friction (or drag) term after $z_d$ in our semi-analytic calculation.  
Therefore we expect 
the dotted line 
is larger than the numerical one as is shown in Figure~\ref{Fig.3}.
After the decoupling epoch, however, the dotted line gradually overlaps with
the numerical one.
We note that this overlapping occurs  earlier 
if we include the drag effect by photons after 
$z_d$ in our free fall calculation.  

On the other hand, if we employ equation (\ref{terminaldb}) by the 
decoupling time $z_d$,  we underestimate the amplitude of baryon
fluctuations.  Therefore the actual ratio between baryon and CDM
density fluctuations is lower than the value of equation (\ref{ratio}).
Nevertheless, baryon density fluctuations have the amplitude 
of $O(10^{-3})\sim O(10^{-2})$ relative to 
CDM fluctuations on small scales at the decoupling epoch.
As we have already mentioned that this ratio does not depend 
on the wavenumber $k$.
These features can be seen in Figure~\ref{Fig.1}.
Since this ratio is fairly small, the assumption 
by HS96, that is to set the baryon density fluctuations to be zero after 
the damping epoch, is verified.
Therefore we can employ their formulas of the evolution of 
CDM density fluctuations 
although some modifications are required on very small scales where 
the wave length of baryon perturbations  is smaller than the Jeans scale
after the decoupling epoch.

\section{Evolution of baryon density fluctuations
after the decoupling}
\def\Ms{{M_\odot}}
\def\yc{{y_C}}
\def\kappamu{{\xi}}
In this section we investigate the evolution of 
density fluctuations
after the decoupling epoch. Baryon density fluctuations are 
catching up with CDM density fluctuations after the decoupling, 
as is shown in Figure~\ref{Fig.1}.
The important physical scale which is relevant to this feature is 
the Jeans scale.
We define the Jeans wavelength (wave number) 
after the decoupling by 
$\lambda_{\rm J}^{\rm phys}={2\pi/k_{\rm J}^{\rm phys}}=
\sqrt{{\pi \cs^2/ G\rho_{\rm m}}}$. Then we have 
$k_{\rm J}=1.3\times10^{3}(1+z)^{1/2}(\gamma\kappamu \Te)^{-1/2}
(\Omega_0 h^2)^{1/2}~\Mpc^{-1} $,
where $\Te$ is the baryon matter temperature in unit of Kelvin, 
and $\kappamu=1-3y_{\rm p}/4$. Correspondingly, the baryon mass 
scale is defined by $M_{\rm J}=
4\pi\rhob(\pi/k^{\rm phys}_{\rm J})^3/3$,
which is written as
\begin{equation}
  \MJ=1.5\times10^{4} (1+z)^{-3/2}(\gamma\kappamu \Te)^{3/2}
  (\Omegam h^2)^{-3/2}\Omegab h^2
  ~\Msolar~.
\end{equation}
Thus the epoch when the Jeans scale becomes smaller than the scale of 
fluctuations with $M$ is 
\begin{equation}
1+z_{\rm J}^{\rm out} = 6.1 \times 10^2 \gamma\kappamu \Te
  (\Omegam h^2)^{-1}(\Omegab h^2)^{2/3}\left(M/M_\odot\right)^{-2/3} ~ .
\end{equation}
As is shown in Figure~\ref{Fig.2}, 
the Jeans scale after the decoupling has the 
plateau. In this stage, the energy transfer between background photons 
and the baryon fluid is effective through the residual electrons, and the
baryon temperature follows the photon temperature.
As the universe expands, however, the energy transfer time rises above 
the Hubble expansion time. After that epoch, 
the baryon matter temperature cools 
adiabatically and drops as $T_{\rm b}\propto a^{-2}$, which derives 
the decrease of the Jeans scale. Thus the Hubble expansion time scale 
and the Compton energy transfer time scale is equal at the broken corner 
of the plateau of the Jeans scale. This epoch is estimated as 
$(1+z)\simeq1000 (\Omegab h^2)^{2/5}$ (\cite{Peebles}).
It is interesting that this redshift depends only on the 
parameter $\Omegab h^2$. 
The line of $\eta_{\rm E}kc_{\rm e}=1$, 
on which the Compton energy time is equal to the sound oscillation time
scale, crosses at the broken corner of the plateau of the Jeans scale.
Thus this cross point of two lines is the point when the sound oscillation
time, the Hubble expansion time and the energy transfer time become all
the same (\cite{YSS}).

This constant value of the Jeans scale,
which is the maximum Jeans scale at the decoupling epoch,
gives a characteristic scale of baryon perturbations.
  From the definition, this 
Jeans scale is written as 
  $k_{\rm JP}=9.0\times10^{2}(\Omega_0 h^2)^{1/2}{\rm Mpc}^{-1}$.
Correspondingly, the baryon mass is 
\begin{equation}
M_{\rm JP }=5.0\times 10^{4}(\Omega_0 h^2)^{-3/2}\Omegab h^2 \Ms ~ .
\end{equation}
Here we have set $y_{\rm p}=0.23$ and $f_{\nu}=0.405$, which we use in the 
hereafter.
According to the usual picture of the Jeans oscillation, 
baryon perturbations whose scale are
smaller than this 
characteristic scale are kept from growing by the baryon thermal pressure.
On the other hand, the density fluctuations 
larger than this characteristic scale can grow by the gravitational
infall into CDM 
potential wells, and catch up to the CDM density fluctuations.
The time scale of this infall process is described by the free fall 
time.

By using the fitting formula,
$T_{\rm b}= 4.5\times 10^{-3}(1+z)^2 (\Omegab h^2)^{-2/5} ~{\rm K}$,
which represents the baryon matter temperature after the energy transfer 
between baryons and photons becomes ineffective 
(\cite{YSS}),
we find that the epoch when the perturbations with the wave 
number $k~(>k_{\rm JP})$ cross the Jeans scale is at redshift
\begin{equation}
1+z_{\rm J}^{\rm out} 
= 2.9\times10^8 (k~ {\rm Mpc})^{-2}\Omega_0 h^2 (\Omegab h^2)^{2/5}.
\end{equation}
This epoch is rewritten in terms of the scale factor as
\begin{equation}
   y_{\rm J}^{\rm out} \equiv 
   a_{\rm J}^{\rm out}/ \aeq = 
   8.2\times 10^{-5} (k~{\rm Mpc})^2 (\Omegab h^2)^{-2/5}.
\label{JeansCross}
\end{equation}
This is the second Jeans crossing time for the perturbations with 
$k>k_{\rm JP}$.   
These perturbations once cross the Jeans scale at $z_{\rm J}$ of 
equation (\ref{jeansin}) before the decoupling epoch
and starts to oscillate as acoustic waves.  
Then again they cross the Jeans scale at $z_{\rm J}^{\rm out}$ and 
are catching up with the CDM density fluctuations
due to gravitational infall.

\section{Transfer Function} 
\def\tq{{\tilde q}}
\def\Omegac{{\Omega_{\rm c}}}
\def\Tm{{T_{\rm m}}}
It will be convenient to recast the evolutionary effects in 
terms of a transfer function of density fluctuations. 
In the present paper, we consider 
the transfer function of  matter density fluctuations
which are defined by 
\begin{equation}
  \delta_{\rm m}={\Omega_{\rm c}\over \Omega_0}\deltac+
  {\Omegab\over \Omega_0}\deltab ~ ,
\label{defofdm}
\end{equation}
with $\Omega_{\rm c}=\Omega_0-\Omegab$. The transfer function 
can be defined by
\begin{equation}
  T_{\rm m}(a,k)={\delta_{\rm m}(a,k)\over 
  \lim_{k\rightarrow 0} \delta_{\rm T }(a,k)} ~ ,
\label{Tm}
\end{equation}
where $\delta_{\rm T}$ is the total transfer function with radiation 
contributions.  In the matter dominate stage,
$\delta_m \simeq \delta_T$.
Using  the large scale solution (HS95), we obtain 
\begin{equation}
  \lim_{k\rightarrow 0} \delta_{\rm m}(a,k)=
  \biggl(1+{4\over15}f_\nu\biggr)
  \biggl(1+{2\over15}f_\nu\biggr)^{-1}
  {6\over5} \biggl({k\over k_{\rm eq}}\biggr)^2 \Phi(0,k) D_1(a),
\end{equation}
where $D_1(a)=2/3+a/a_{\rm eq}$.   
In general, the transfer function is a function of time.
However, once the pressure term of baryon perturbations 
becomes negligible at some scale,
$\delta_{\rm m}\propto D_1(a)$ (Peebles 1980;~see also equation (\ref{D1sol})).
Then the transfer function (\ref{Tm}) becomes independent of time at
this scale.
Therefore we can assume the time invariance of the transfer function 
after the decoupling epoch on scales larger than the maximum Jeans
scale after the decoupling, i.e., $k < k_{\rm JP}$ or after 
$z_{\rm J}^{\rm out}$ on $k > k_{\rm JP}$.

\begin{figure}[t]
\centerline{\epsfxsize=10cm \epsffile{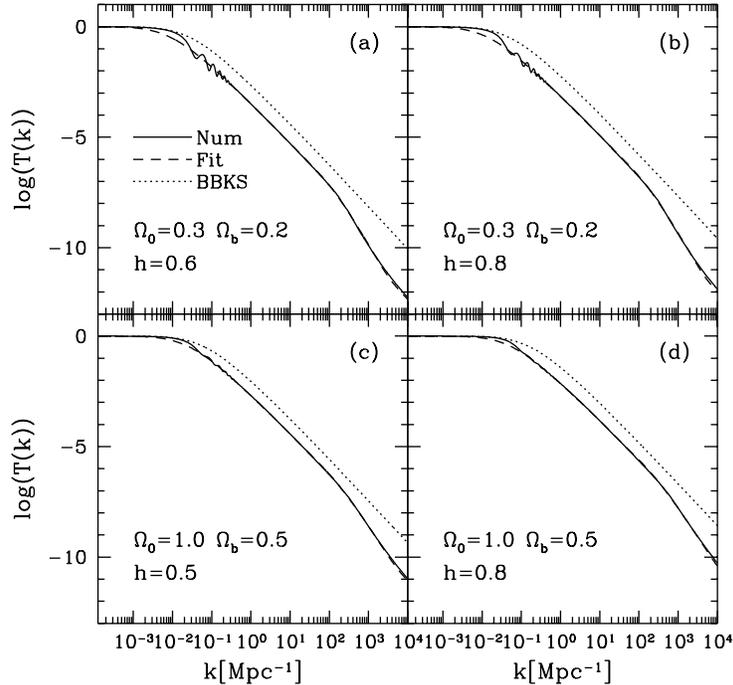}}
\caption{\protect\footnotesize
 Transfer functions $T_{\rm m}(k)$ at present.
 They are obtained by Numerical calculations (solid lines), analytic fitting 
 (dashed lines) and BBKS fitting (dotted lines), respectively.
}
\label{Fig.4}
\end{figure}
While the transfer function by Bardeen, et al. (1986, hereafter BBKS) 
is familiar,
the effect of the baryon fraction is not taken into account.
This transfer function is not applicable for models with non-negligible
baryon fraction at $k\gtrsim 0.1 \Omegab^{1/2} \Omega_0^{1/2} h^2 
{\rm Mpc}^{-1}$.
The effects of the baryon fraction on the large-scale power spectrum 
have been investigated by Holtzman (1989), Peacock \& Dodds (1995), 
and Sugiyama~(1995). The effects of the baryon 
fraction on the transfer function is incorporated by simply 
scaling of so-called $\Gamma$ factor of the BBKS transfer function
\footnote{In a recent paper by Eisenstein
  \& Hu (1997),  they obtained better scaling, which includes $k$
  dependence, to fit intermediate scale ($\sim
  100$Mpc).  Moreover, they presented very
  detailed and precise, but complicated,  
  fitting formula of the transfer function on 
  large scales ($k \lesssim 1 \rm Mpc^{-1})$ with taking into account 
  the acoustic oscillations of the photon-baryon fluid before the
  decoupling epoch.  
  Although their paper
  appears after writing up this paper, we decide to mention their 
  transfer function in the following sessions.  
}.
However, errors arise on scales smaller than the galaxy scale.
In Figure~\ref{Fig.4}, the fitting formula of the transfer function  
and the numerical computation described below 
are compared with the BBKS transfer function. 
Here the deviation is exaggerated by considering
the cosmological models with high baryon density fraction.
HS96 have investigated the CDM transfer functions 
on very small scales. Their result is summarized in the first part of 
Appendix B.
In their paper, the effect of the baryon fraction 
is taken into account and they have given the  transfer function 
in an analytic form. Though it is rather complicated equation,
it has a high accuracy even for high baryon density models.
However, there are following limitations in their transfer function.
   First, their transfer function is limited only on very small scales 
($k \gtrsim 1\rm Mpc$), 
and it is not applicable at the scales larger than the galaxy scale.
Secondly, they do not take into account the Jeans oscillation of baryon
fluctuations after the decoupling epoch.  Therefore their transfer function 
cannot apply at the wave number larger than $k_{\rm JP}$.
Here we propose an analytic transfer function which is 
useful for the entire region of wavelength.

The transfer function which we propose is following:
\begin{equation}
  T_{\rm m}(a,k)={{\rm ln}\bigl(1+2.34\tq(1+\kappa\tq)/(1+\tq)\bigr) 
  \over 
  2.34\tq\bigl[1+3.89\tq+(16.1\tq)^2+(5.46\tq)^3+(6.71\tq)^4\bigr]^{1/4}}
  f_{\rm m}(a,k),
\label{TmA}
\end{equation}
with
\begin{eqnarray}
  &&\tq=0.951\bigl(\alpha\Omegac/\Omega_0\bigr)^{-1/2}
  {k {\rm Mpc}\over 
  \Omega_0 h^2 
} 
  \Theta_{2.7}^2~,
\\
  &&\kappa=0.809\bigl(\alpha\Omegac/\Omega_0\bigr)^{1/2} \beta,
\end{eqnarray}
where $\Theta_{2.7}=T_0/2.7{\rm K}$, 
$\alpha$ and $\beta$ are defined in equation (\ref{alphabeta}), 
and $f_{\rm m}(a,k)$ in equation (\ref{fmaappen}) in Appendix B. 
Except for $f_{\rm m}(a,k)$, the transfer function is obtained
by incorporating the small scale transfer function in HS96 
into the familiar formula of BBKS with scaling. The
large scale behavior of the transfer function at 
$k\sim 0.1{\rm Mpc}^{-1}$ is successfully expressed
by BBKS transfer function with simple scaling (\cite{Sugiyama}).
However, the small scale behavior is not satisfactory.
It is required that the transfer function approaches to 
the form in HS96 in the small scale limit.
The term in the logarithm in equation (\ref{TmA}) 
plays a role to change from large scale formula to
the small scale one.

Let us briefly comment on the correction factor $f_{\rm m}(a,k)$,
which is described in detail in Appendix B.
As discussed in \S 4, the maximum Jeans scale after the decoupling epoch 
is expressed by $k_{\rm JP}$. The baryon thermal pressure 
keeps baryon fluctuations from growing even after the decoupling epoch
on the scales $k>k_{\rm JP}$.
When baryon fluctuations can not grow and remain the small 
amplitude, the growth rate of CDM density fluctuations is suppressed. 
In this stage, the growth rate of CDM fluctuations is roughly 
in proportion to $(1+a/\aeq)^{-\alpha_1}$ with 
$\alpha_1=(1-\sqrt{1+24(\Omega_{\rm 0}-\Omega_{\rm b})/\Omega_0})/4$.
Eventually this effect decreases the transfer function on sales 
smaller than $k_{\rm JP}$ 
after the decoupling.  
In order to describe this effect, we introduce the window function 
$f_{\rm m}(a,k)$.

In Figure~5, the fitting formula of the transfer function  
is compared with the numerical computation,  normalized
to the BBKS transfer function. 
Our fitting formula reproduces the numerical result 
quite well for the scale, $1 {\rm Mpc}^{-1}\lesssim k 
\lesssim 100{\rm Mpc}^{-1}$, within a few percent for 
$0.5\lesssim h\lesssim 0.8$ and $\Omegab/\Omega_0\lesssim 0.5$. 
   For the smaller scales $k\gtrsim100{\rm Mpc}^{-1}$,
where $f_{\rm m}(a,k)$ becomes important, our fitting
function works within $\sim 10 \%$ accuracy at $z\lesssim100$.
  For the larger scales $k\lesssim 0.1{\rm Mpc}^{-1}$, the numerical 
transfer 
function shows the feature with bumps and wriggles 
due to the baryon-photon acoustic oscillation
before the decoupling. The larger the baryon fraction is, 
the bigger this effect becomes.
As we do not take this effect into account, the difference
between the fitting formula and the numerical 
result becomes large. 
However, the fitting formula still 
designs to cross the center of these oscillations.
Therefore we expect fairly good fit with the observational quantities,
such as $\sigma_8$.
The accuracy of the fitting formula
is checked in the next section in more detail by computing
statistical quantities.

\vspace{1cm}

\centerline{
\makebox[\textwidth][t]{\vbox{\parindent=0pt \baselineskip=12pt
\hbox{{\epsfxsize=8cm \epsffile{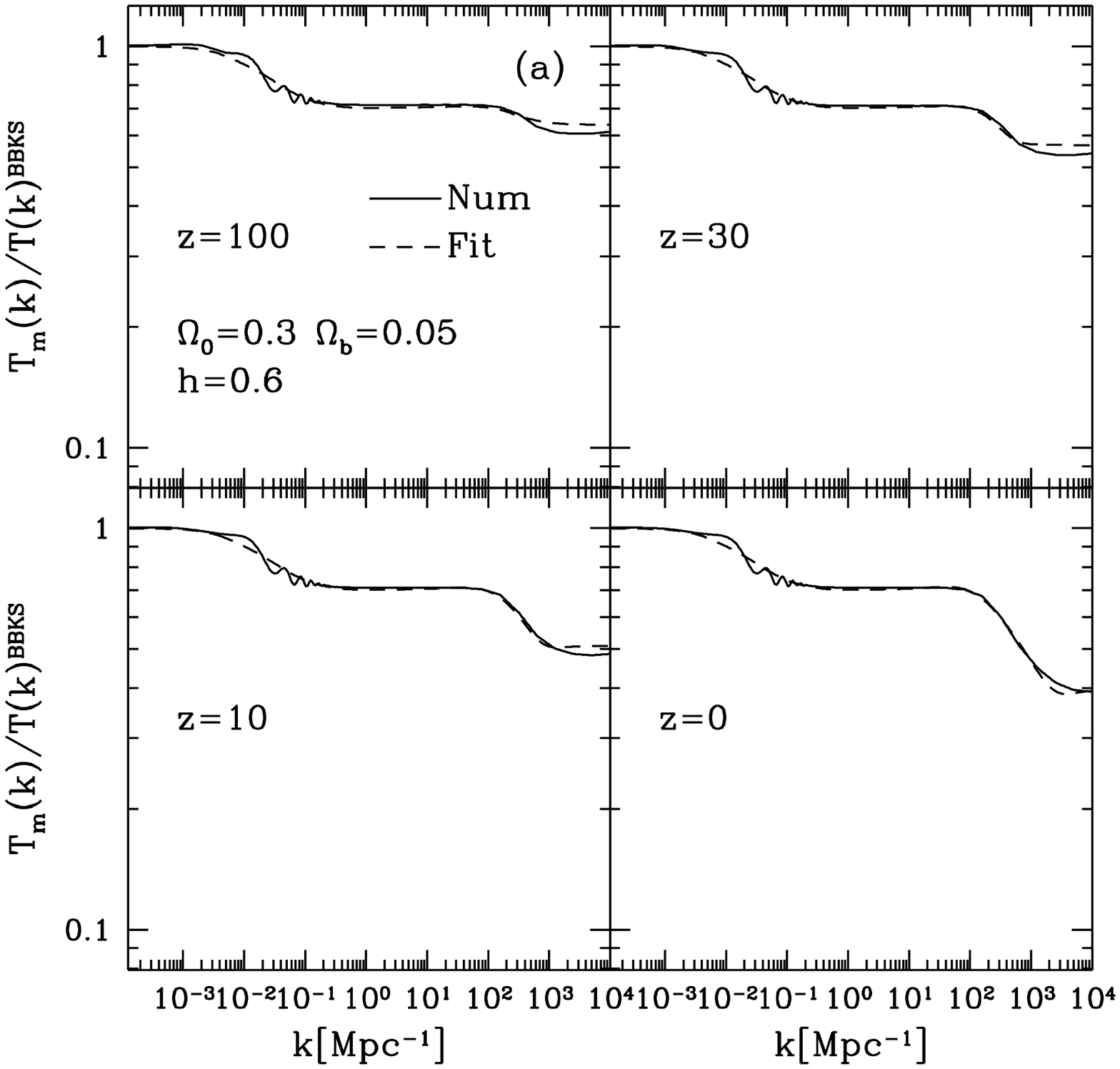}} \hspace{0.5cm}
    {\epsfxsize=8cm
    \epsffile{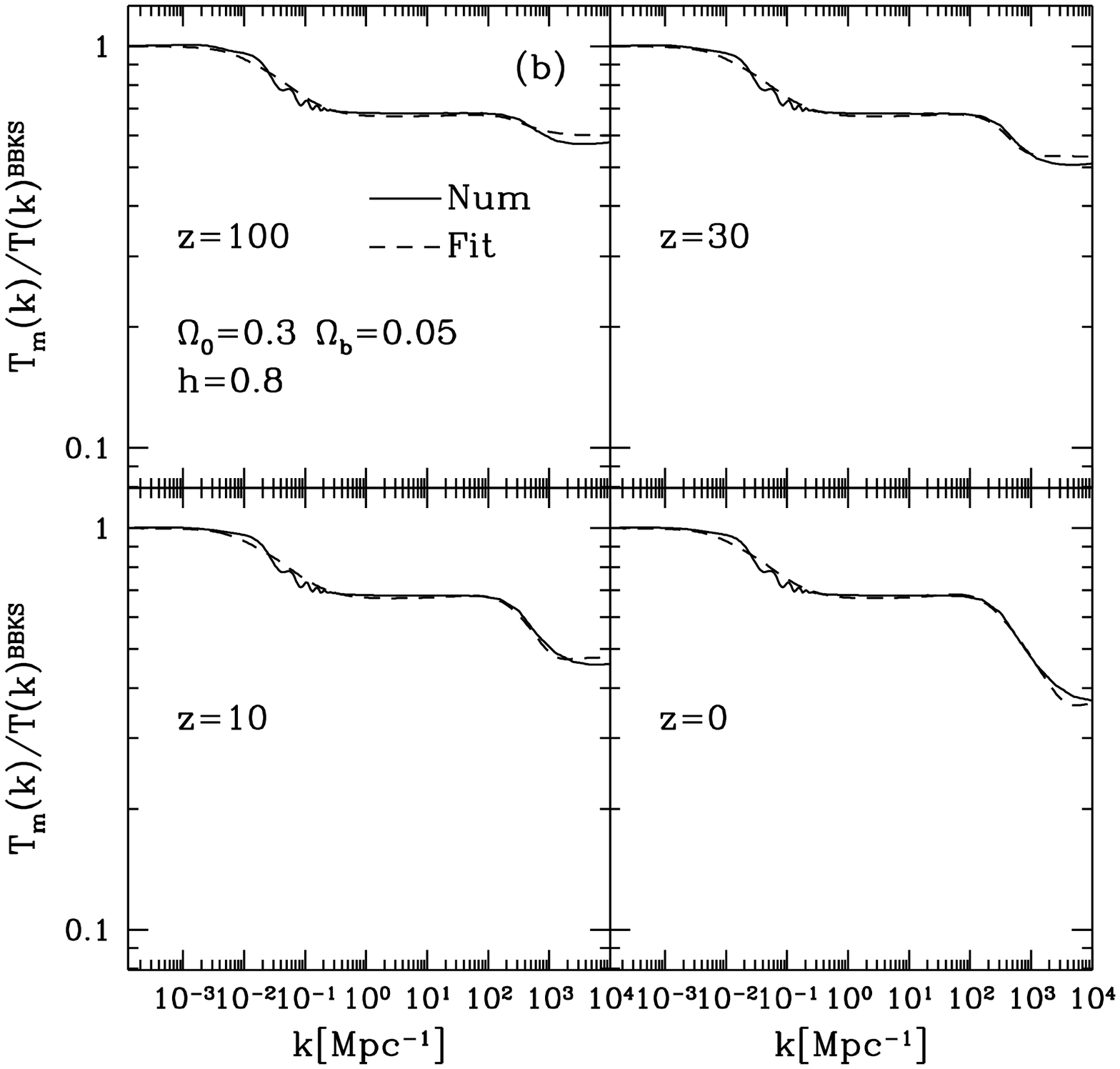}}}
\hbox{{\epsfxsize=8cm \epsffile{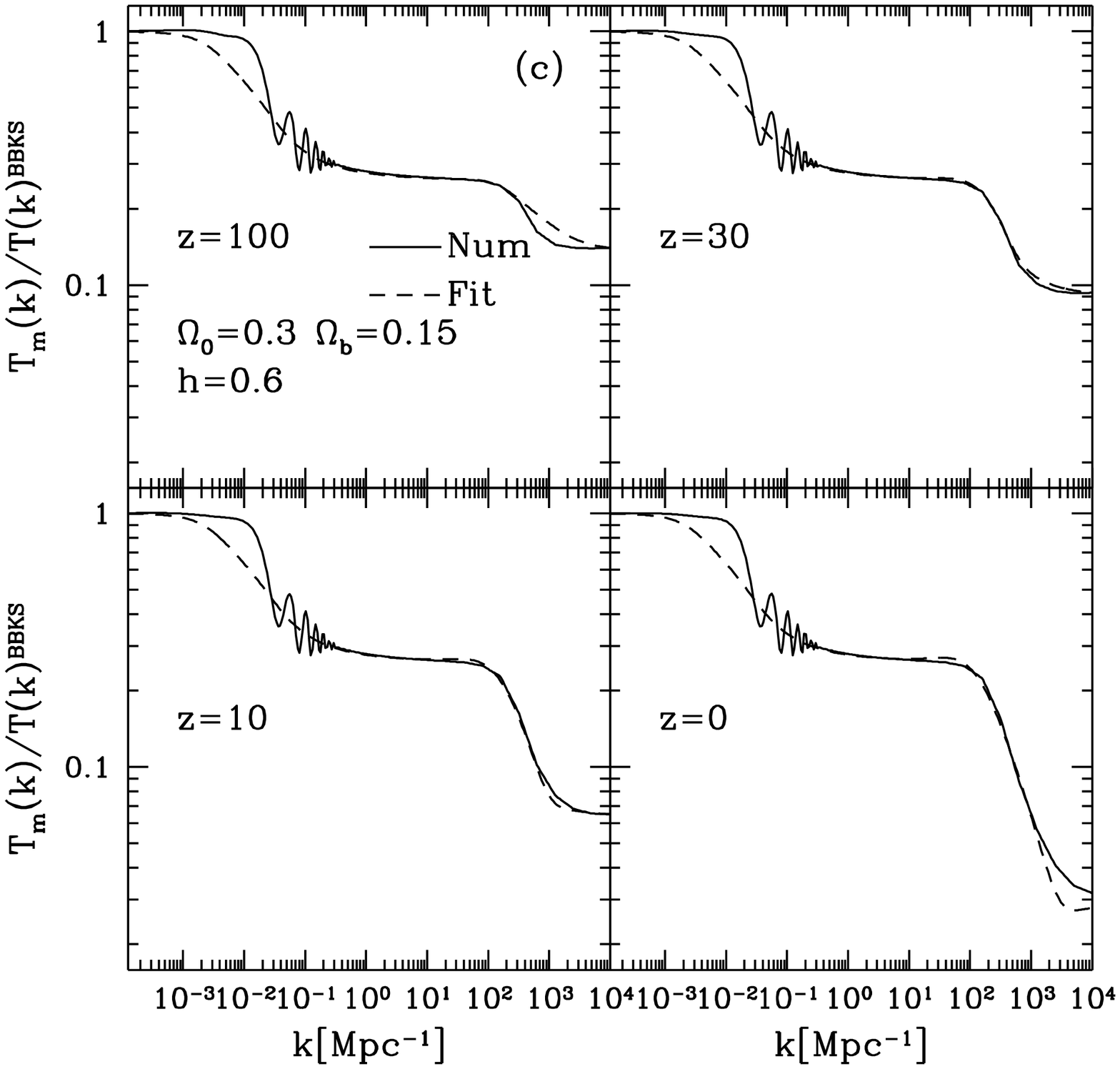}}\hspace{0.62cm}
    {\epsfxsize=8cm
    \epsffile{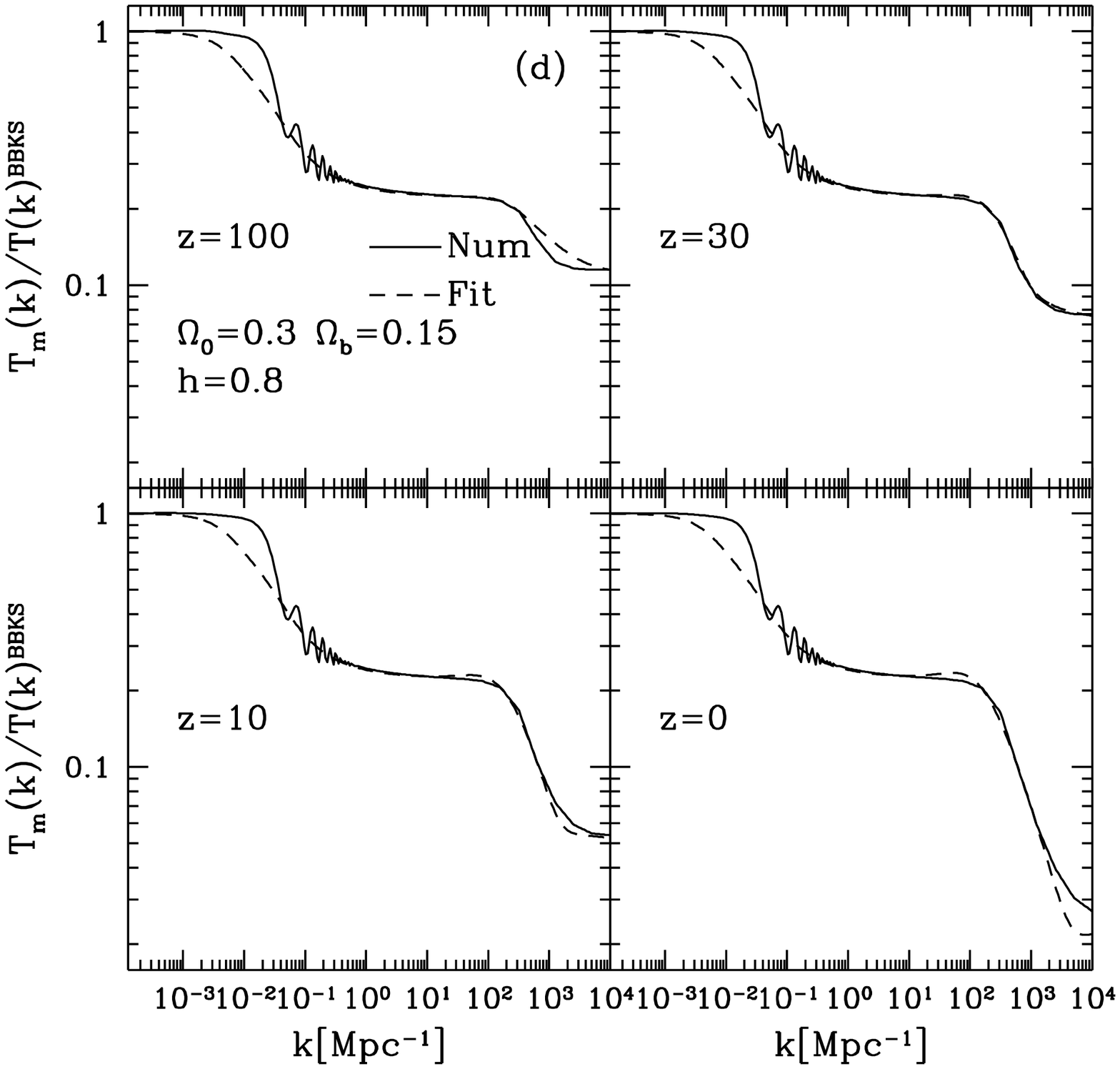}}}
\vspace{0.5cm}
\vbox{{\noindent 
Fig.5.---{\protect\footnotesize 
Transfer functions normalized by the BBKS transfer function
$T_{\rm m}(k)/T^{\rm BBKS}(k)$. In each panel, the fitting formula
is compared with the numerical result. 
4 panels in each figure  show the time evolution of the transfer function 
at $z=100$, $z=30$, $z=10$, $z=0$, respectively. 
}}}
}}
}
\newpage

\centerline{
\makebox[\textwidth][t]{\vbox{\parindent=0pt \baselineskip=12pt
\hbox{{\epsfxsize=8cm \epsffile{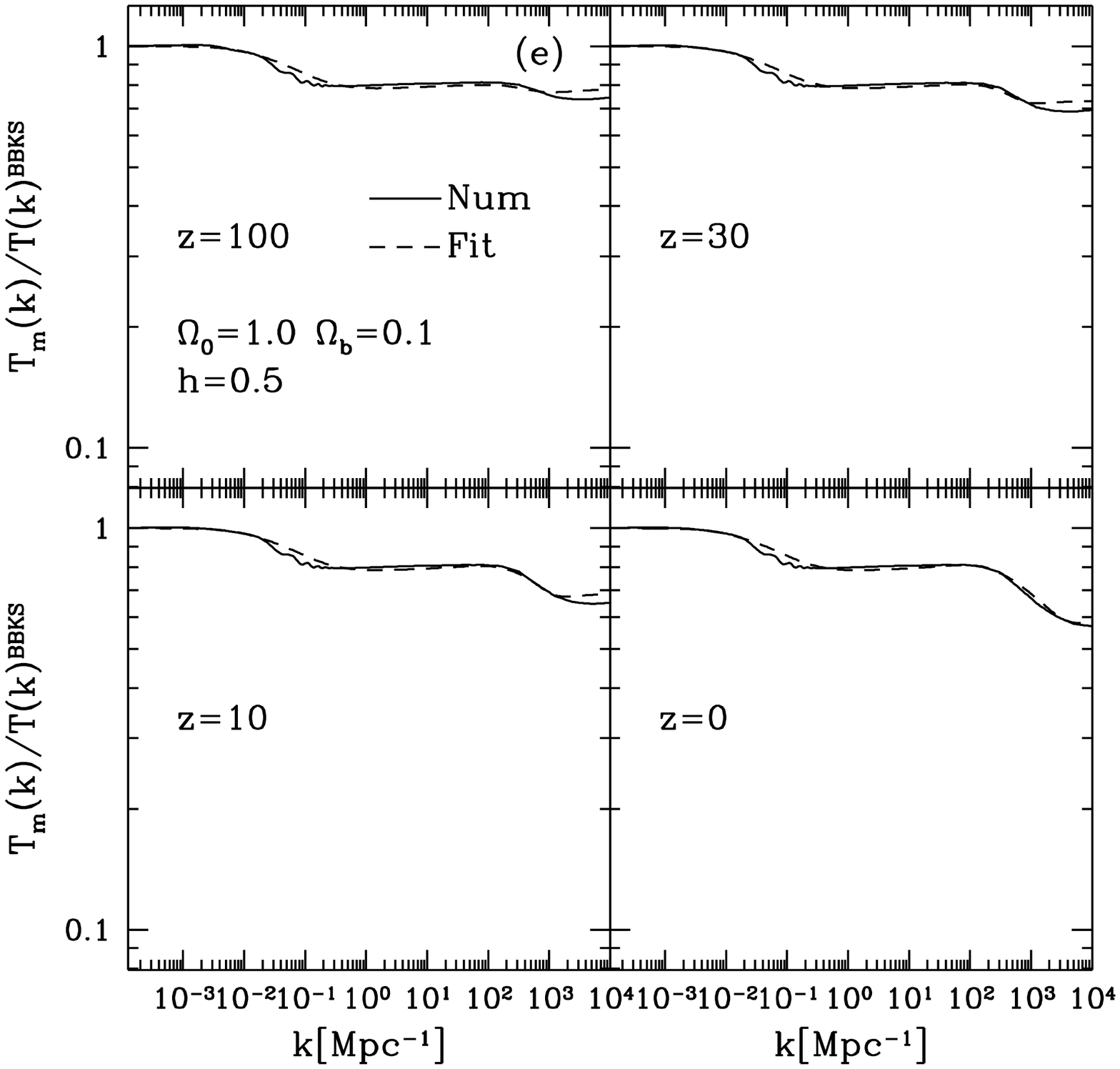}} \hspace{0.5cm}
    {\epsfxsize=8cm
    \epsffile{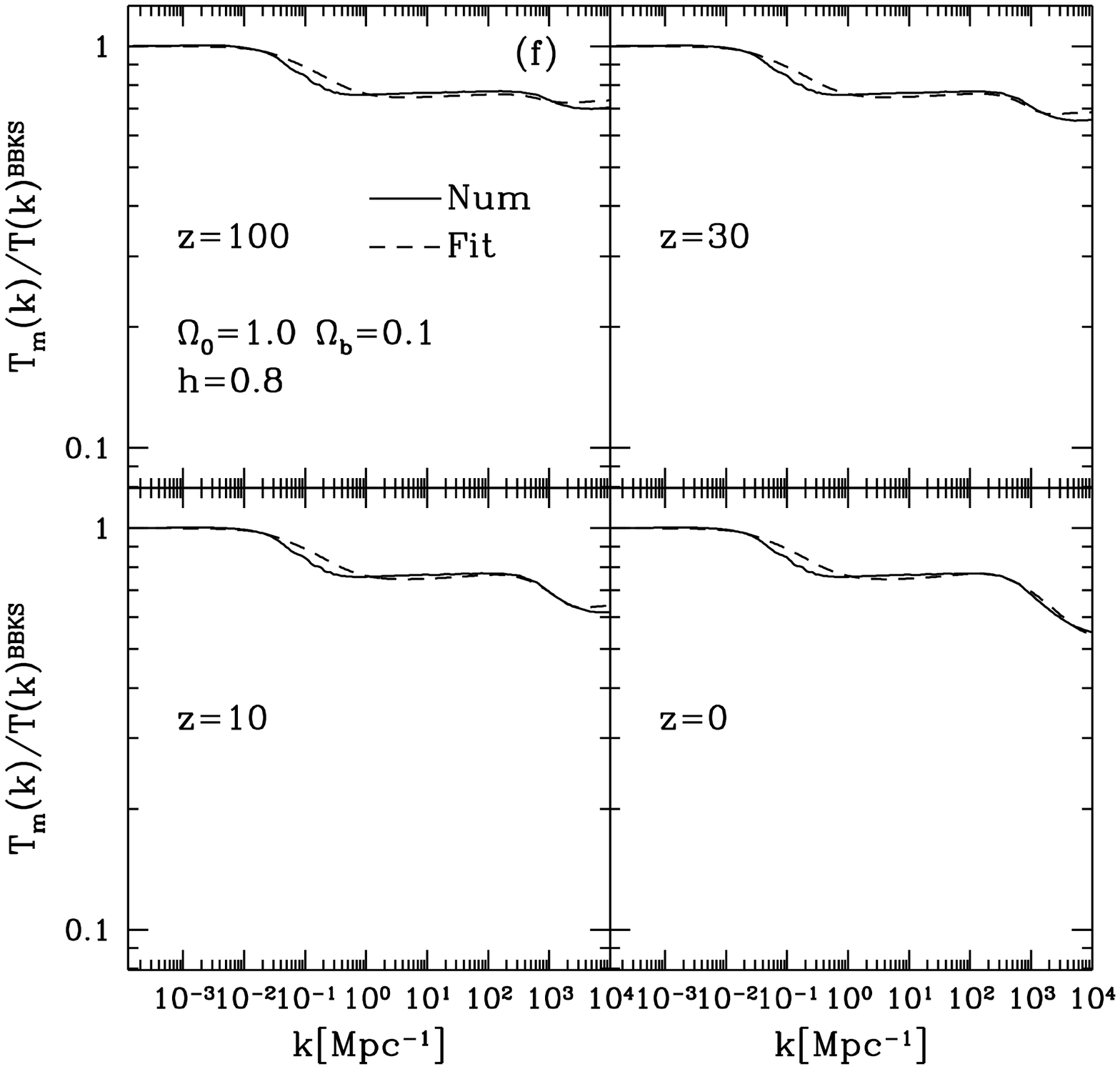}}}
\hbox{{\epsfxsize=8cm \epsffile{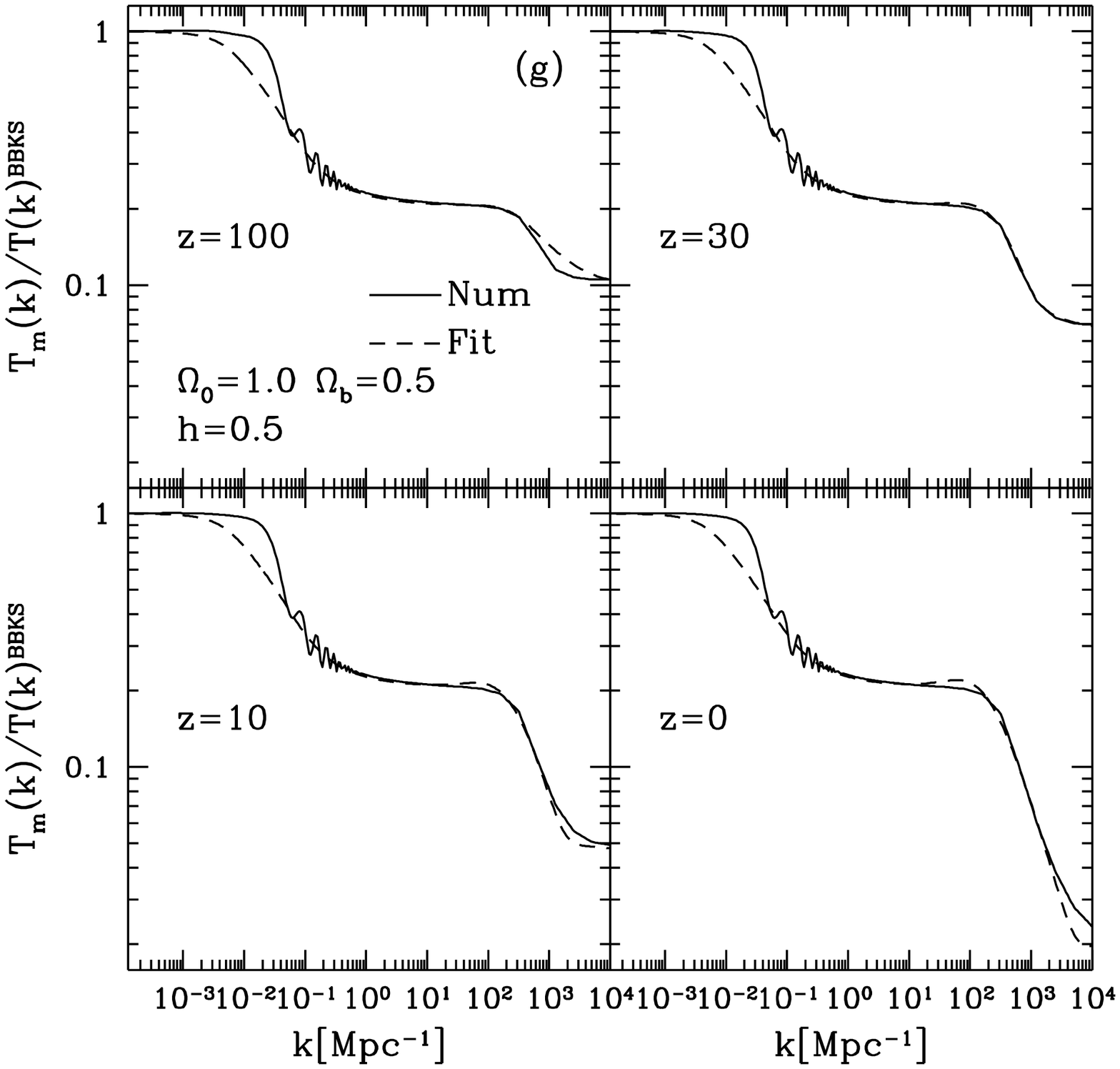}}\hspace{0.62cm}
    {\epsfxsize=8cm
    \epsffile{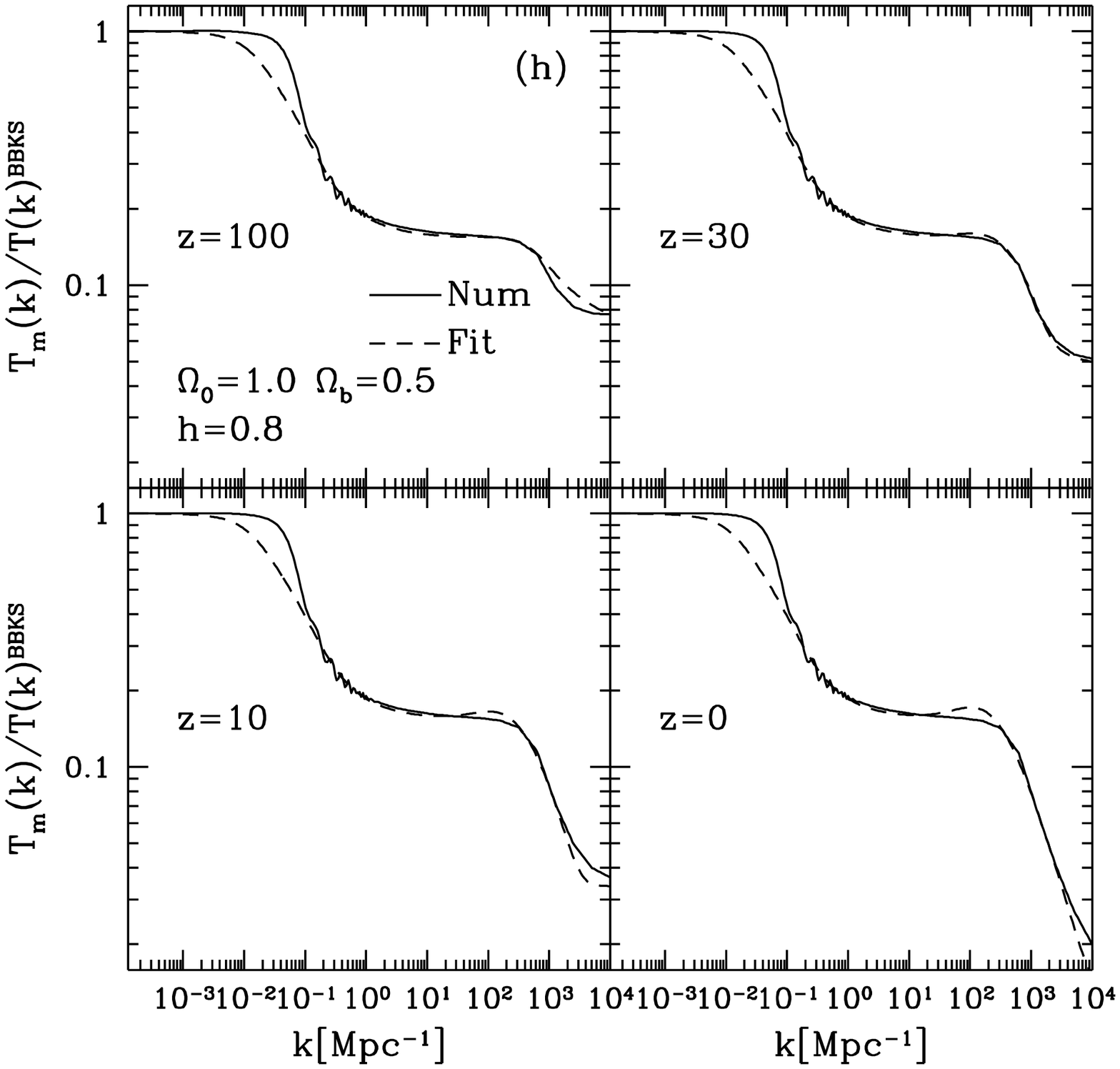}}}
\vspace{0.5cm}
\vbox{{\noindent 
Fig.5.---{\protect\footnotesize 
Continued}}}
}}
}
\newpage


\section{Simple Demonstrations}
In this section, we first show the accuracy of the fitting formula
by calculating the following statistical quantities of the large scale
structure the universe.
Here we consider the CDM cosmology 
with the Harrison-Zeldovich initial density power spectrum.
    For low density universe models, we introduce the cosmological constant
to keep the flat geometry.  
Let us introduce $\sigma_R$ which is defined as 
\begin{equation}
  \sigma^2_R(a)={1\over 2\pi^2}\int_0^\infty dk k^2 P(a,k) W(kR)^2 ~ ,
\label{sigma}
\end{equation}
where $P(a,k)=\bigl<\deltam(a,k)^2\bigr>$ is the power spectrum of 
the linear density perturbations at time $a$,  
and $W(x)=3(\sin x-x\cos x)/x^3$ is the top-hat window 
function. To determine the  amplitude of the
power spectrum, we use the COBE normalization by Bunn \& White (1996). 
According to their results, we can rewrite equation
(\ref{sigma}) as
\begin{equation}
  \sigma^2_R(a)=\biggl({D_1(a)\over D_1(a_0)}\biggr)^2
  \delta_H^2\int_0^\infty {dk\over k}
  \biggl({k\over H_0}\biggr)^4 T_{\rm m}(a,k)^2 W(kR)^2,
\label{sigmadef}
\end{equation}
with $\delta_H=1.94\times10^{-5} \Omega_0^{-0.785-0.05{\rm ln}\Omega_0}$
for $\Lambda$-models. We note that $D_1(a)$ in equation (\ref{sigmadef}) 
represents the growing mode solution for matter perturbations 
taking $\Lambda$-term into account, which differs from (\ref{D1sol}) 
at $z\lesssim$ a few due to the $\Lambda$-term.
   For $\Lambda$-model, $D_1(a)$ is written as (see e.g., Peebles 1980)
\begin{equation}
  D_1(a(1+z))={5\Omega_0\over2}
  \sqrt{{\Omega_0 (1+z)^3}+1-\Omega_0}
  \int_0^{1/(1+z)} da'
  \biggl({a'\over\Omega_0+a'^3(1-\Omega_0)}\biggr)^{3/2}~,
\end{equation}
which is normalized as $D_1(a)\simeq 1/(1+z)$ at $z\gg1$.

The observational quantity $\sigma_8$ is 
defined as $\sigma_8
=\sigma_{R=8h^{-1}{\rm Mpc}}(z=0)$. 
In Table~\ref{tab:A}, we show the value of $\sigma_8$ 
by numerical calculations
comparing with various transfer functions.
In this table, $\sigma_8^{\rm N}$,
$\sigma_8^{\rm S}$, $\sigma_8^{\rm F}$, $\sigma_8^{\rm BBKS}$ 
and $\sigma_8^{\rm EH}$
denote values obtained by the numerical calculation, 
our fitting formula, the empirical scaling by Sugiyama (1995),
the original fitting formula by BBKS and 
the scaling\footnote{
Here we used $\Gamma_{\rm eff}(k)$ in their paper 
in stead of the usual $\Gamma$-factor $(=\Omega_0h)$.} 
by Eisenstein \& Hu (1997), respectively.

It is shown that our fitting formula together with that obtained
by the empirical scalings by the shape parameter $\Gamma$ 
by (\cite{Sugiyama})  and by 
Eisenstein \& Hu (1997),
well reproduce the numerical result. This is because that the fitting 
formulas are designed to cross the center of oscillations.

In order to check the accuracy on small scales,
we have calculated the quantities,
$\sigma_{R=1h^{-1}{\rm Mpc}}(z=3)$,
$\sigma_{R=0.1h^{-1}{\rm Mpc}}(z=5)$, 
$\sigma_{R=10^{-2}h^{-1}{\rm Mpc}}(z=10)$,  and
$\sigma_{R=10^{-3}h^{-1}{\rm Mpc}}(z=10)$,
for various cosmological models.
The results are shown in Tables~\ref{tab:B}-\ref{tab:E}. 
    From these results, we can find that our fitting formula works 
very  well on small scales.

Next we briefly demonstrate a cosmological implication of our results.
The study of the early formation of collapsed objects and the thermal 
history of the high-$z$ universe is
one of the most important issue
in cosmology and galaxy formation.
(e.g., \cite{FK}; \cite{GO}; \cite{OG}; Haiman \& Loeb 1997).
The statistical arguments allow us to investigate
such early history of the formation of small scale 
cosmic objects in the high-$z$ universe
without big numerical calculations.
The Press-Schechter theory (Press \& Schechter 1974)
is among the most simple theory to discuss 
the statistics of the gravitationally collapsing objects.
According to the Press-Schechter formula,
we calculate the fractions of the gravitationally
bound system, and show the effects of the baryon fraction
on the formation rate of the nonlinear objects in the high-$z$ 
universe.

\begin{figure}[t]
\centerline{\epsfxsize=10cm \epsffile{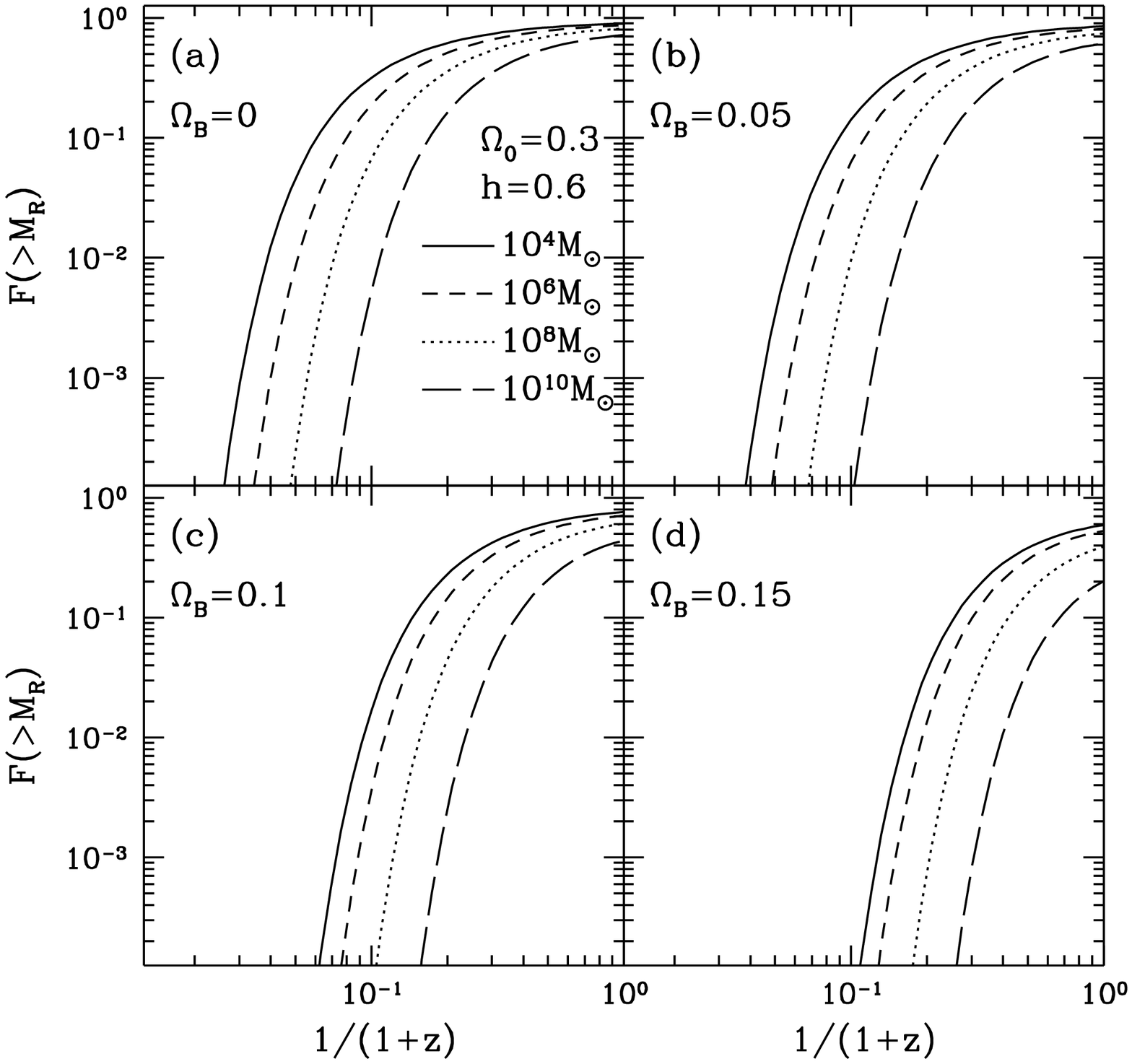}}
\figurenum{6}
\caption{\protect\footnotesize
  Mass fractions of the bound objects $F(>M_R)$.
  The cosmological parameters taken in this figure is
  $\Omega_0=0.3$ and $h=0.6$. In each panel  the baryon
  fraction is different, which are:
  (a)$\Omegab=0$;
  (b)$\Omegab=0.05$;
  (c)$\Omegab=0.1$;
  and (d)$\Omegab=0.15$. 
}
\label{Fig.6}
\end{figure}
The Press-Schechter theory predicts that 
the mass fraction which is gravitationally bounded objects above a 
given mass $M_R$ at cosmic time $a$  becomes (e.g., \cite{Padmanabhan}),
\begin{equation}
  F(>M_R)={\rm erfc}\biggl({\bar\delta_{c}\over \sqrt{2}
\sigma (M_R,a)}\biggr),
\label{erfc}
\end{equation}
where $\sigma(M_R,a) \equiv \sigma_R(a)$ 
is define by equation (\ref{sigmadef}),
$M_R=4\pi\bar\rho_{\rm 0} R^3/3$, 
$\bar\rho_{\rm 0}$ is the spatially averaged matter energy density,
and $\bar\deltac=1.69$.
    Figure~\ref{Fig.6} shows the mass fraction $F(>M_R)$ calculated from 
equation (\ref{erfc}), for the cosmological model $\Omega_0=0.3$, $h=0.6$
with different baryon density fractions.
(a) is the case that the baryon fraction is neglected.
(b) is the case with including the baryon fraction $\Omegab=0.05$.
(c) is the case $\Omegab=0.1$, and (d) is $\Omegab=0.15$.
It is apparent that the formation rate of 
cosmic objects is delayed in a universe with large baryon fraction.

\section{Summary and Discussions}
In this paper, we have carefully re-investigated the evolution of the 
small scale density fluctuations in an expanding universe, 
motivated from the studies of cosmic structure formation in the
high-$z$ universe. 
Under the assumption of the hierarchical clustering scenario, 
the structure formation takes place from  smaller scales.  
Therefore such careful investigation of 
the evolution of density fluctuations on small scales is 
necessary.
In the universe with non-negligible baryon fraction, the evolution of 
density fluctuations is very complicated because 
baryon
perturbations affect CDM perturbations through the gravitational 
interaction.

Interestingly, baryon density fluctuations show the varieties of 
physical processes caused by
the interaction with photons which are summarized 
in Figure~\ref{Fig.2}.
Discussing the physical scales characterizing these processes, 
we have studied the evolution of the small scale baryon 
density perturbations.
In this investigation, we have found the 
growth of baryon density fluctuations on small scales
after the diffusion damping and before the decoupling epoch, 
which is related with the 
breaking down of the tight coupling approximation 
between baryons and photons.   We have referred 
this growing mode as the terminal velocity mode.

We have re-analyzed the total matter transfer function.
In the universe with non-negligible baryon fraction,
the transfer function is changed from the 
familiar transfer function by BBKS,
for $k\gtrsim 0.1 \Omega_0{}^{1/2}\Omega_0{}^{1/2} h^2$.
Extending the previous fitting formulas by Sugiyama (1995)
and HS96 with including the Jeans effect after the decoupling
epoch which causes the damping on very small scales, 
we have presented the analytic 
transfer function in a way of modifying the BBKS transfer function.
This fitting function is rather complicated, but its accuracy
is very high particularly around the galaxy scale 
$1{\rm Mpc}^{-1}\lesssim k \lesssim 100{\rm Mpc}^{-1}$. 
This fitting function is applicable for the entire range.
Even for $k\gtrsim100 {\rm Mpc}^{-1}$  it works within  
$\sim 10 \%$  error.
Unless one would like to trace the oscillation feature  on 
intermediate scales 
$ 0.01{\rm Mpc}^{-1} \lesssim k \lesssim 0.1{\rm Mpc}^{-1}$ 
which is caused by the acoustic 
oscillation by photon-baryon fluid before the decoupling epoch
(in this case, one need to employ pure
numerical calculations or the newly suggested fitting formula by 
Eisenstein \& Hu (1997)), 
our fitting works fairly well for the entire range of 
$k\lesssim 10^4{\rm Mpc}^{-1}$.

When discussing the formation epoch of small cosmic objects,
the amplitude of density fluctuations is the important factor.
In the universe with large baryon density fraction, the amplitude 
of total matter perturbations is decreased on small scales. 
Then the formation of the cosmic objects is delayed. 
Moreover, the evolution of total density fluctuations on these scales 
does not quite follow the usual growth rate $D_1$.  We have described this 
time dependence of the transfer function as $f_{\rm m}$.
In a low density universe, the baryon density fraction 
becomes large relative to the CDM density fraction. 
Then this effect becomes significant, as demonstrated in \S 6.

    Finally, if once the early re-ionization occurs, 
baryon density perturbations
evolve in a different way.  This effect is beyond the scope of 
this paper.  
But a recent paper by Chiba, Kawasaki \& Sugiyama (1997) showed 
that the baryon and total matter evolution are affected by 
the early re-ionization in case of high baryon fraction models.

\vspace{0.3cm}
\begin{center}
{\bf ACKNOWLEDGMENT}
\end{center}

One of us (K.Y.) thanks Y.Kojima and N.Ohuo for comments and discussions
at the early stage of this work. He also thanks J.Yokoyama and the
people at Yukawa Institute and Department of Physics, Kyoto University, 
Kitashirakawa, Kyoto, where some parts of this work were done, 
for their hospitality.
This work is supported by the Grants-in-Aid for 
Scientific Research of Ministry of Education, Science
and Culture of Japan (Nos.~09740203 and 09440106).

\vspace{0.3mm}

\newpage 

\vspace{10mm}
\begin{appendix}
\section{ Evolution of Matter Perturbations }
In this Appendix, we briefly review the analytic formulation 
of the evolution of the matter perturbations on small scales
obtained by HS96. 
Let us consider the evolution after the diffusion damping. 
If we set photon and baryon fluctuations to be negligible small
(this assumption is valid by the argument in \S 3.3 even though 
there exists small 
fraction of baryon fluctuations after the damping epoch due to the
terminal velocity), 
only CDM contributes to the gravitational potential.
Which permits  the equation of CDM perturbations to be written as 
\begin{equation}
  {d^2 \deltac\over dy^2}+{2+3y\over 2y(1+y)}{d\deltac\over dy}=
  {3\over 2y(1+y)}{\Omega_{\rm c}\over \Omega_0}\deltac,
\end{equation}
where $y\equiv a/\aeq$ and $\Omega_{\rm c}=\Omega_0-\Omegab$.
The solutions are given in terms of the Hypergeometric function
\begin{equation}
  U_i=(1+y)^{-\alpha_i} 
  {}_2F_1\Bigl(\alpha_i,\alpha_i+1/2,2\alpha_i+1/2;1/(1+y)\Bigr)~,
\label{HypU}
\end{equation}
where $i=1,2$ and $\alpha_i=(1\pm\sqrt{1+24\Omega_{\rm c}/\Omega_0})/4$
with $-$ and $+$ for $i=1$ and $2$, respectively.

With considering the matching conditions to the solution
outside the horizon in the radiation dominated stage,
the solution for $\delta_{\rm c}$ after the horizon crossing
is obtained, 
\begin{equation}
  \delta_{\rm c}(\eta,k)=I_{1}\Phi(0,k) 
  \Bigl( A_1 U_1(\eta)+A_2 U_2(\eta)\Bigr),
\label{deltacan}
\end{equation}
where
\begin{equation}
  A_1={\Gamma(\alpha_1)\Gamma(\alpha_1+1/2)
  \over \Gamma(2\alpha_1+1/2)(\psi(2\alpha_2)-\psi(2\alpha_1)) }
  \biggl({1\over2}\ln \Bigl[I_2{a_{\rm eq}\over a_{\rm H}}\Bigr]
  +\psi(1)-\psi(\alpha_2)+\ln2\biggr),
\end{equation}
$A_2$ is obtained by replacing the subscripts $1\leftrightarrow2$,
$I_1$ and $I_2$ are obtained from the numerical fitting formulas
\begin{eqnarray}
  &&I_1=9.11(1+0.128\fnu+0.029\fnu^2),
\\
  &&I_2=0.594(1-0.631\fnu+0.284\fnu^2),
\\
  &&{a_{\rm H}\over a_{\rm eq}}={1+\sqrt{1+8(k/\keq)^2}\over4(k/\keq)^2},
\end{eqnarray}
and $\psi(x)$ is the poli-gamma function.

After the epoch that the baryon  pressure 
becomes negligible, i.e., the drag epoch $z_{\rm d}$ 
or the second Jeans crossing epoch $z_{\rm J}^{\rm out}$ depends on scales, 
baryon fluctuations evolve through CDM gravitational potential well. 
In this stage, it is well known  that the matter fluctuations 
$\delta_{\rm m}$, defined by equation (\ref{defofdm}), 
follow the growing and decaying solutions (Peebles 1980) 
\begin{eqnarray}
  &&D_1={2\over 3}+y,
\label{D1sol}
\\
  &&D_2={15\over 8}(2+3y) {\rm ln}
  \biggl[ {(1+y)^{1/2}+1\over (1+y)^{1/2}-1}\biggl]
  -{45\over 4}(1+y)^{1/2}.
\label{D2sol}
\end{eqnarray}
These solutions correspond to $U_1$ and $U_2$ with 
$\alpha_i \vert_{\Omega_c=\Omega_0} = (1 \pm 5 )/4$ . 
Matching solutions at $y\equiv y_{\rm c} = y_{\rm d}$ or 
$= y_{\rm J}^{\rm out}$, we obtain for the growing 
mode solution as 
\begin{equation}
  \delta_{\rm m}(\eta,k)=
  \biggl( G_1\delta_{\rm m}-G_2\dot\delta_{\rm m} \biggr)
  \bigg\vert_{y=y_{\rm c}} D_1(y(\eta)),
\label{freefall}
\end{equation}
with
\begin{equation}
  G_1={\dot D_2\over D_1\dot D_2-\dot D_1 D_2~},
  \hspace{7mm}
  G_2={     D_2\over D_1\dot D_2-\dot D_1 D_2}~.
\end{equation}

\section{ Effect of the baryon thermal pressure on the Transfer Function}
In this Appendix we describe some details which are important
to construct an analytic transfer function on very small scales, 
$k > 1 {\rm Mpc}^{-1}$. 
We first note the small scale transfer function obtained in
HS96. According to their result, 
the fitting formula of the CDM transfer function is written as,
\begin{equation}
  T_{\rm c}(a,k)=\alpha {{\rm ln} \bigl[1.8 \beta q\bigr]\over 14.2 q^2},
\end{equation}
with
\begin{equation}
  \alpha=a_1^{-\Omegab/\Omega_0}a_2^{-(\Omegab/\Omega_0)^3},
\hspace{7mm}
  \beta^{-1}=1+b_1\bigr((\Omega_{\rm c}/\Omega_0)^{b_2}-1\bigr),
\label{alphabeta}
\end{equation}
\begin{eqnarray}
  &&a_1=(46.9\Omega_0h^2)^{0.670}
    \bigl(1+(32.1\Omega_0h^2)^{-0.532}\bigr), 
\nonumber
\\
  &&a_2=(12.0\Omega_0h^2)^{0.424}
     \bigl(1+(45.0\Omega_0h^2)^{-0.582}\bigr),
\nonumber
\\
  &&b_1=0.944\bigl(1+(458.0\Omega_0h^2)^{-0.708}\bigr)^{-1},
\nonumber
\\
  &&b_2=(0.395\Omega_0h^2)^{-0.0266},
\nonumber
\end{eqnarray}
and $q=k({\rm Mpc^{-1}} \Omega_0 h^2)^{-1}(T_0/2.7{\rm K})^2$.
Though this formula is rather complicated, it works to $1 \%$ accuracy.
We should notice that this formula is applicable only on small scales,
$k \gtrsim  1 {\rm Mpc}^{-1}$. 

Let us describe the function $f_{\rm m}(a,k)$ in equation (\ref{TmA}). 
As we have shown in Appendix A, $\deltam$ grows in proportion to
$D_1(a)$ after the pressure becomes ineffective. 
On scales larger than the Jeans scale at the decoupling epoch, i.e., 
$k < k_{\rm JP}$, 
the photon pressure becomes ineffective after the decoupling epoch.
Then $\deltam$ evolves in proportion to $D_1(a)$.
Once the matter fluctuations become to follow this growing mode 
solution, the transfer function becomes independent of time.
Therefore on these scales, the transfer function does not change its
shape after the drag  epoch.
On scales $k\gtrsim k_{\rm JP}$,
on the other hand, the baryon thermal pressure keeps the perturbations
from growing even after the decoupling of baryons from photons. 
As we have seen in Appendix A, the growth rate of CDM 
fluctuations is suppressed  in the situation that there exists 
a homogeneous matter (baryon) component besides the CDM component.
This effect alters the transfer function on small scales
$k\gtrsim k_{\rm JP}$ even after the decoupling epoch.
The function $f_{\rm m}(a,k)$ is multiplied in order to
describe this effect.

In order to find the analytic form of $f_{\rm m}(a,k)$, we rewrite
\begin{equation}
  \Tm(a,k)=\Tm(a_{\rm d},k){\Tm(a,k)\over \Tm(a_{\rm d},k)}
  \equiv \Tm(a_{\rm d},k) f(a,k),
\end{equation}
where $a_{\rm d}$ denotes the scale factor at the decoupling epoch.
Thus $f(a,k)$ is defined as the ratio of the transfer 
function at $a(\eta)$ to that at the decoupling time.
From the definition of the transfer function (\ref{Tm}), 
we can rewrite 
\begin{equation}
  f(a,k)={D_1(a_{\rm d})\over D_1(a)}
  {\deltam(a,k)\over \deltam(a_{\rm d},k)}.
\end{equation}

We first consider the case $k<k_{\rm JP}$. 
This case is trivial, because $\deltam$ grows in proportion to 
$D_1(a)$ after the decoupling, as described before. 
Then we get $f(a,k)=1$.

We next consider the case $k>k_{\rm JP}$. These perturbations stay 
inside the Jeans scale until 
$a_{\rm J}^{\rm out}/\aeq=y_{\rm J}^{\rm out}$, which is
defined in equation~(\ref{JeansCross}). 
Assuming that baryon density fluctuations do 
not contribute to the gravitational potential in this regime, 
we approximate 
the growth rate of the CDM density perturbations by 
the growing mode solution $U_1(a)$ in 
equation~(\ref{HypU}). 
After the decoupling epoch, we can assume the matter domination of the
universe, i.e., 
$a/\aeq\gg1$.  Hence we further 
approximate $U_1(a)\simeq (1+a/\aeq)^{\alpha_1}$ with  
$\alpha_1=(1 - \sqrt{1+24\Omega_{\rm c}/\Omega_0})/4$ as is shown in 
Appendix A.
Then we can write
\begin{equation}
  f(a,k)
  \simeq{D_1(a_{\rm d})\over D_1(a)}{U_1(a)\over U_1(a_{\rm d})}
  \simeq\biggl({1+y_{\rm d}\over 1+y}\biggr)^{1+\alpha_1},
\hspace{2cm}
  {(y <y_{\rm J}^{\rm out})}, 
\end{equation}
where $y = a/\aeq$ and $y_{\rm d} = a_{\rm d}/a$ as are defined before.
After the Jeans crossing $a/\aeq>y_{\rm J}^{\rm out}$, we approximate that 
$\deltam$ follows the growing mode $D_1(a)$ and 
that the solution is given by equation (\ref{freefall})
with replacing $y_{\rm c}$ by $y_{\rm J}^{\rm out}$. 
Then we can write
\begin{eqnarray}
  f(a,k)
  &\simeq&{D_1(a_{\rm d})\over U_1(a_{\rm d})}
  \biggl(G_1(a) U_1(a)-G_2(a) \dot U_1(a)\biggr)\bigg\vert_{a/\aeq=y_{\rm J}^{\rm out}}
\nonumber
\\
  &\simeq&
  {3-2\alpha_1\over 5}
  \biggl({1+a_{\rm d}/\aeq\over 1+y_{\rm J}^{\rm out}}\biggr)^{1+\alpha_1},
\hspace{3cm}
  {(a/\aeq> y_{\rm J}^{\rm out})}.
\end{eqnarray}
Here we have assumed that the amplitude of $\deltab$ is small
at $y_{\rm J}^{\rm out}$.

Summarizing the results, $f(a,k)$ is evaluated
\[
  f(a,k)\simeq\left\{
    \begin{array}{ll}
       1 &\hspace{2cm}(k<k_{\rm JP}),
\\
       \Bigl({(1+y_{\rm d})/ (1+y)}\Bigr)^{\alpha_1+1} 
       &\hspace{2cm}(k>k_{\rm JP} ~{\rm and} ~y<y_{\rm J}^{\rm out}),
\\
       \Bigl({(1+y_{\rm d})/( 1+y_{\rm J}^{\rm out})}\Bigr)^{\alpha_1+1}
      (3-2\alpha_1)/5 
       &\hspace{2cm}(k>k_{\rm JP} ~{\rm and} ~y>y_{\rm J}^{\rm out}).
    \end{array}
  \right.
\]
In the case when the baryon fraction is negligible, 
i.e., $\Omega_{\rm b}\ll \Omega_0$, we have $\alpha_1=-1$ and $f(a,k)=1$.
On the other hand, in the cosmological models with the 
non-negligible baryon 
fraction, the growth rate of CDM fluctuations
is suppressed even after the decoupling. 
And the matter transfer function is changed.

Combining above $f(a,k)$'s on difference scales, 
we give the next fitting function,
\begin{eqnarray}
  &&f_{\rm m}(a,k)={1\over (k/\tilde k_{\rm JP})^{1.3}+1}
   +{1\over( \tilde k_{\rm JP}/k)^{1.3}+0.9}
   \biggl[{1\over (y/y_{\rm J}^{\rm out})^{1.3}+1}
   \biggl({1+y_{\rm d}\over 1+y}\biggr)^{\alpha_1+1}
\nonumber
\\  &&\hspace{6.5cm} 
  +{1\over (y_{\rm J}^{\rm out}/y)^{1.3}+1}{3-2\alpha_1\over5}
   \biggl({1+y_{\rm d}\over 1+y_{\rm J}^{\rm out}}\biggr)^{\alpha_1+1}
   \biggl],
\label{fmaappen}
\end{eqnarray}
with
\begin{eqnarray}
  &&\tilde k_{\rm JP}={350 ( 1+50~(y_{\rm d}/y))}
  (\Omega_0 h^2)^{1/2} {\rm Mpc}^{-1},
\\
  &&y_{\rm d}=20 \Omega_0 h^2 (\Omegab h^2)^{-0.034},
\end{eqnarray}
and $y=a/\aeq=2.4\times10^4 (1+z)^{-1}\Omega_0 h^2$.
Here we determine numerical constant factors to reproduce the 
numerical computations.
The reason why we employ $\tilde k_{\rm JP}$ instead of $k_{\rm JP}$
is following.
Even on scales smaller than  $k_{\rm JP}$, 
the thermal pressure cannot entirely prevent baryon fluctuations from
the growth because of the existence of the CDM potential.
A certain amount of the growth rate has to be taken into account at
$k \gtrsim k_{\rm JP}$ even though the rate is less than $D_1$.

\end{appendix}


\newpage
\begin{deluxetable}{cccccccc}
\footnotesize
\tablecaption{$\sigma_{R=8 h^{-1} {\rm Mpc}}$ for various cosmological 
  models.  \label{tab:A}}
\tablewidth{0pt}
\tablehead{
\colhead{$\Omega_0$} &\colhead{$\Omegab$} &\colhead{$h$} &
\colhead{$\sigma_{8 h^{-1}{\rm Mpc}}^{{\rm N}}$} &
\colhead{$\sigma_{8 h^{-1}{\rm Mpc}}^{{\rm F}}$} &
\colhead{$\sigma_{8 h^{-1}{\rm Mpc}}^{{\rm S}}$} &
\colhead{$\sigma_{8 h^{-1}{\rm Mpc}}^{{\rm BBKS}}$} &
\colhead{$\sigma_{8 h^{-1}{\rm Mpc}}^{{\rm EH}}$}}
\startdata
$0.3$ & $0.05$ & $0.6$  & $0.70$ & $0.71$ & $0.70$ & $0.95$ & $0.71$ \nl 
$0.3$ & $0.1 $ & $0.6$  & $0.51$ & $0.51$ & $0.51$ & $0.95$ & $0.50$ \nl 
$0.3$ & $0.15$ & $0.6$  & $0.35$ & $0.33$ & $0.37$ & $0.95$ & $0.34$ \nl 
$0.3$ & $0.2 $ & $0.6$  & $0.23$ & $0.19$ & $0.26$ & $0.95$ & $0.22$ \nl 
$0.3$ & $0.05$ & $0.8$  & $0.98$ & $1.00$ & $0.98$ & $1.36$ & $0.99$ \nl 
$0.3$ & $0.1 $ & $0.8$  & $0.68$ & $0.69$ & $0.70$ & $1.36$ & $0.68$ \nl 
$0.3$ & $0.15$ & $0.8$  & $0.46$ & $0.44$ & $0.49$ & $1.36$ & $0.44$ \nl 
$0.3$ & $0.2 $ & $0.8$  & $0.29$ & $0.24$ & $0.34$ & $1.36$ & $0.28$ \nl 
$1.0$ & $0.05$ & $0.5$  & $1.18$ & $1.22$ & $1.18$ & $1.31$ & $1.21$ \nl 
$1.0$ & $0.1 $ & $0.5$  & $1.08$ & $1.12$ & $1.06$ & $1.31$ & $1.11$ \nl 
$1.0$ & $0.5 $ & $0.5$  & $0.50$ & $0.46$ & $0.41$ & $1.31$ & $0.48$ \nl 
$1.0$ & $0.05$ & $0.8$  & $1.83$ & $1.92$ & $1.87$ & $2.07$ & $1.90$ \nl 
$1.0$ & $0.1 $ & $0.8$  & $1.65$ & $1.76$ & $1.68$ & $2.07$ & $1.73$ \nl 
$1.0$ & $0.5 $ & $0.8$  & $0.74$ & $0.66$ & $0.62$ & $2.07$ & $0.70$ \nl 
\enddata
\tablenotetext{a}{$\sigma^{\rm N}, \sigma^{\rm F}, \sigma^{\rm S}, 
\sigma^{\rm BBKS}$, and 
$\sigma^{\rm EH}$ are calculated by the full numerical computation, 
using our analytic fitting, an analytic fitting by Sugiyama (1995), the 
one by BBKS and 
the one by Eisenstein \& Hu (1997) with employing their empirical
scalings of the shape parameter $\Gamma$, respectively.}
\end{deluxetable}

\begin{deluxetable}{cccccccc}
\footnotesize
\tablecaption{$\sigma_{R=1h^{-1}{\rm Mpc}}$ at $z=3$ \label{tab:B}}
\tablewidth{0pt}
\tablehead{
\colhead{$\Omega_0$} &\colhead{$\Omegab$} &\colhead{$h$} &
\colhead{$\sigma_{1 h^{-1}{\rm Mpc}}^{{\rm N}}$} &
\colhead{$\sigma_{1 h^{-1}{\rm Mpc}}^{{\rm F}}$} &
\colhead{$\sigma_{1 h^{-1}{\rm Mpc}}^{{\rm S}}$} &
\colhead{$\sigma_{1 h^{-1}{\rm Mpc}}^{{\rm BBKS}} $} &
\colhead{$\sigma_{1 h^{-1}{\rm Mpc}}^{{\rm EH}}$} }
\startdata
  $0.3$  &  $0.05$  & $0.6$  &  $0.66$  &  $0.66$  &  $0.64$  &  $0.92$ & $0.65$ \nl 
  $0.3$  &  $0.1$  &  $0.6$  &  $0.44$  &  $0.44$  &  $0.44$  &  $0.92$ & $0.43$ \nl 
  $0.3$  &  $0.15$  & $0.6$  &  $0.28$  &  $0.27$  &  $0.30$  &  $0.92$ & $0.26$ \nl 
  $0.3$  &  $0.2$  &  $0.6$  &  $0.16$  &  $0.15$  &  $0.20$  &  $0.92$ & $0.14$ \nl 
  $0.3$  &  $0.05$  & $0.8$  &  $1.00$  &  $0.99$  &  $0.96$  &  $1.45$ & $0.97$ \nl 
  $0.3$  &  $0.1$  &  $0.8$  &  $0.64$  &  $0.63$  &  $0.64$  &  $1.45$ & $0.61$ \nl 
  $0.3$  &  $0.15$  & $0.8$  &  $0.38$  &  $0.37$  &  $0.42$  &  $1.45$ & $0.36$ \nl 
  $0.3$  &  $0.2$  &  $0.8$  &  $0.20$  &  $0.19$  &  $0.27$  &  $1.45$ & $0.19$ \nl 
  $1.0$  &  $0.05$  & $0.5$  &  $1.31$  &  $1.31$  &  $1.25$  &  $1.45$ & $1.29$ \nl 
  $1.0$  &  $0.1$  &  $0.5$  &  $1.16$  &  $1.16$  &  $1.08$  &  $1.45$ & $1.15$ \nl 
  $1.0$  &  $0.5$  &  $0.5$  &  $0.37$  &  $0.37$  &  $0.31$  &  $1.45$ & $0.36$ \nl 
  $1.0$  &  $0.05$  & $0.8$  &  $2.51$  &  $2.53$  &  $2.44$  &  $2.86$ & $2.51$ \nl 
  $1.0$  &  $0.1$  &  $0.8$  &  $2.18$  &  $2.21$  &  $2.07$  &  $2.86$ & $2.18$ \nl 
  $1.0$  &  $0.5$  &  $0.8$  &  $0.58$  &  $0.57$  &  $0.53$  &  $2.86$ & $0.56$ \nl 
\enddata
\tablenotetext{a}{$\sigma^{\rm N}, \sigma^{\rm F}, \sigma^{\rm S}, 
\sigma^{\rm BBKS}$, and 
$\sigma^{\rm EH}$ are same as Table 1.}
\end{deluxetable}
\newpage
\begin{deluxetable}{cccccccc}
\footnotesize
\tablecaption{$\sigma_{R=0.1h^{-1}{\rm Mpc}}$ at $z=5$ 
 \label{tab:C}}
\tablewidth{0pt}
\tablehead{
\colhead{$\Omega_0$} &\colhead{$\Omegab$} &\colhead{$h$} &
\colhead{$\sigma_{0.1 h^{-1}{\rm Mpc}}^{{\rm N}}$} &
\colhead{$\sigma_{0.1 h^{-1}{\rm Mpc}}^{{\rm F}}$} &
\colhead{$\sigma_{0.1 h^{-1}{\rm Mpc}}^{{\rm S}}$} &
\colhead{$\sigma_{0.1 h^{-1}{\rm Mpc}}^{{\rm BBKS}}$} &
\colhead{$\sigma_{0.1 h^{-1}{\rm Mpc}}^{{\rm EH}}$} 
}
\startdata
 $0.3$  &  $0.05$  & $0.6$  &  $0.94$  &  $0.93$  &  $0.88$  &  $1.32$ & $0.89$ \nl 
 $0.3$  &  $0.1$  &  $0.6$  &  $0.61$  &  $0.60$  &  $0.59$  &  $1.32$ & $0.57$ \nl 
 $0.3$  &  $0.15$  & $0.6$  &  $0.36$  &  $0.36$  &  $0.39$  &  $1.32$ & $0.33$ \nl 
 $0.3$  &  $0.2$  &  $0.6$  &  $0.19$  &  $0.19$  &  $0.26$  &  $1.32$ & $0.17$ \nl 
 $0.3$  &  $0.05$  & $0.8$  &  $1.47$  &  $1.46$  &  $1.38$  &  $2.16$ & $1.39$ \nl 
 $0.3$  &  $0.1$  &  $0.8$  &  $0.91$  &  $0.90$  &  $0.88$  &  $2.16$ & $0.84$ \nl 
 $0.3$  &  $0.15$  & $0.8$  &  $0.51$  &  $0.51$  &  $0.56$  &  $2.16$ & $0.47$ \nl 
 $0.3$  &  $0.2$  &  $0.8$  &  $0.25$  &  $0.25$  &  $0.35$  &  $2.16$ & $0.24$ \nl 
 $1.0$  &  $0.05$  & $0.5$  &  $2.26$  &  $2.23$  &  $2.08$  &  $2.47$ & $2.16$ \nl 
 $1.0$  &  $0.1$  &  $0.5$  &  $1.98$  &  $1.95$  &  $1.76$  &  $2.47$ & $1.89$ \nl 
 $1.0$  &  $0.5$  &  $0.5$  &  $0.55$  &  $0.55$  &  $0.45$  &  $2.47$ & $0.51$ \nl 
 $1.0$  &  $0.05$  & $0.8$  &  $4.83$  &  $4.75$  &  $4.48$  &  $5.41$ & $4.63$ \nl 
 $1.0$  &  $0.1$  &  $0.8$  &  $4.13$  &  $4.06$  &  $3.71$  &  $5.41$ & $3.93$ \nl 
 $1.0$  &  $0.5$  &  $0.8$  &  $0.93$  &  $0.91$  &  $0.80$  &  $5.41$ & $0.84$ \nl 
\enddata
\tablenotetext{a}{$\sigma^{\rm N}, \sigma^{\rm F}, \sigma^{\rm S}, 
\sigma^{\rm BBKS}$, and 
$\sigma^{\rm EH}$ are same as Table 1.}
\end{deluxetable}

\begin{deluxetable}{cccccccc}
\footnotesize
\tablecaption{$\sigma_{R=10^{-2}h^{-1}{\rm Mpc}}$ at $z=10$  
 \label{tab:D}}
\tablewidth{0pt}
\tablehead{
\colhead{$\Omega_0$} &\colhead{$\Omegab$} &\colhead{$h$} &
\colhead{$\sigma_{10^{-2} h^{-1}{\rm Mpc}}^{{\rm N}}$} &
\colhead{$\sigma_{10^{-2} h^{-1}{\rm Mpc}}^{{\rm F}}$} &
\colhead{$\sigma_{10^{-2} h^{-1}{\rm Mpc}}^{{\rm S}}$} &
\colhead{$\sigma_{10^{-2} h^{-1}{\rm Mpc}}^{{\rm BBKS}}$} &
\colhead{$\sigma_{10^{-2} h^{-1}{\rm Mpc}}^{{\rm EH}}$} 
}
\startdata
  $0.3$  &  $0.05$  & $0.6$  &  $0.85$  &  $0.85$  &  $0.79$  &  $1.20$ & $0.80$ \nl 
  $0.3$  &  $0.1$  &  $0.6$  &  $0.54$  &  $0.54$  &  $0.52$  &  $1.20$ & $0.50$ \nl 
  $0.3$  &  $0.15$  & $0.6$  &  $0.32$  &  $0.32$  &  $0.34$  &  $1.20$ & $0.29$ \nl 
  $0.3$  &  $0.2$  &  $0.6$  &  $0.16$  &  $0.16$  &  $0.22$  &  $1.20$ & $0.14$ \nl 
  $0.3$  &  $0.05$  & $0.8$  &  $1.37$  &  $1.36$  &  $1.26$  &  $2.02$ & $1.27$ \nl 
  $0.3$  &  $0.1$  &  $0.8$  &  $0.82$  &  $0.83$  &  $0.79$  &  $2.02$ & $0.75$ \nl 
  $0.3$  &  $0.15$  & $0.8$  &  $0.45$  &  $0.46$  &  $0.49$  &  $2.02$ & $0.41$ \nl 
  $0.3$  &  $0.2$  &  $0.8$  &  $0.22$  &  $0.22$  &  $0.31$  &  $2.02$ & $0.20$ \nl 
  $1.0$  &  $0.05$  & $0.5$  &  $2.28$  &  $2.25$  &  $2.05$  &  $2.46$ & $2.14$ \nl 
  $1.0$  &  $0.1$  &  $0.5$  &  $1.98$  &  $1.96$  &  $1.72$  &  $2.46$ & $1.85$ \nl 
  $1.0$  &  $0.5$  &  $0.5$  &  $0.52$  &  $0.53$  &  $0.41$  &  $2.46$ & $0.47$ \nl 
  $1.0$  &  $0.05$  & $0.8$  &  $5.12$  &  $5.04$  &  $4.63$  &  $5.65$ & $4.79$ \nl 
  $1.0$  &  $0.1$  &  $0.8$  &  $4.34$  &  $4.28$  &  $3.79$  &  $5.65$ & $4.03$ \nl 
  $1.0$  &  $0.5$  &  $0.8$  &  $0.91$  &  $0.93$  &  $0.75$  &  $5.65$ & $0.79$ \nl 
\enddata
\tablenotetext{a}{$\sigma^{\rm N}, \sigma^{\rm F}, \sigma^{\rm S}, 
\sigma^{\rm BBKS}$, and 
$\sigma^{\rm EH}$ are same as Table 1.}
\end{deluxetable}

\newpage
\begin{deluxetable}{cccccccc}
\footnotesize
\tablecaption{$\sigma_{R=10^{-3}h^{-1}{\rm Mpc}}$ at $z=10$  
 \label{tab:E}}
\tablewidth{0pt}
\tablehead{
\colhead{$\Omega_0$} &\colhead{$\Omegab$} &\colhead{$h$} &
\colhead{$\sigma_{10^{-3} h^{-1}{\rm Mpc}}^{{\rm N}}$} &
\colhead{$\sigma_{10^{-3} h^{-1}{\rm Mpc}}^{{\rm F}}$} &
\colhead{$\sigma_{10^{-3} h^{-1}{\rm Mpc}}^{{\rm S}}$} &
\colhead{$\sigma_{10^{-3} h^{-1}{\rm Mpc}}^{{\rm BBKS}}$} &
\colhead{$\sigma_{10^{-3} h^{-1}{\rm Mpc}}^{{\rm EH}}$} 
}
\startdata
 $0.3$  &  $0.05$ &  $0.6$  &  $1.16$  &  $1.15$  &  $1.15$  &  $1.77$ & $1.17$ \nl 
 $0.3$  &  $0.1$  &  $0.6$  &  $0.68$  &  $0.68$  &  $0.75$  &  $1.77$ & $0.72$ \nl 
 $0.3$  &  $0.15$ &  $0.6$  &  $0.38$  &  $0.38$  &  $0.49$  &  $1.77$ & $0.41$ \nl 
 $0.3$  &  $0.2$  &  $0.6$  &  $0.19$  &  $0.19$  &  $0.32$  &  $1.77$ & $0.20$ \nl 
 $0.3$  &  $0.05$ &  $0.8$  &  $1.88$  &  $1.86$  &  $1.86$  &  $3.02$ & $1.88$ \nl 
 $0.3$  &  $0.1$  &  $0.8$  &  $1.05$  &  $1.04$  &  $1.15$  &  $3.02$ & $1.10$ \nl 
 $0.3$  &  $0.15$  & $0.8$  &  $0.55$  &  $0.55$  &  $0.71$  &  $3.02$ & $0.59$ \nl 
 $0.3$  &  $0.2$  &  $0.8$  &  $0.25$  &  $0.26$  &  $0.44$  &  $3.02$ & $0.29$ \nl 
 $1.0$  &  $0.05$  & $0.5$  &  $3.48$  &  $3.47$  &  $3.16$  &  $3.80$ & $3.29$ \nl 
 $1.0$  &  $0.1$  &  $0.5$  &  $2.97$  &  $2.94$  &  $2.63$  &  $3.80$ & $2.83$ \nl 
 $1.0$  &  $0.5$  &  $0.5$  &  $0.69$  &  $0.70$  &  $0.60$  &  $3.80$ & $0.69$ \nl 
 $1.0$  &  $0.05$  & $0.8$  &  $8.05$  &  $8.00$  &  $7.30$  &  $8.97$ & $7.56$ \nl 
 $1.0$  &  $0.1$  &  $0.8$  &  $6.68$  &  $6.62$  &  $5.93$  &  $8.97$ & $6.32$ \nl 
 $1.0$  &  $0.5$  &  $0.8$  &  $1.23$  &  $1.25$  &  $1.12$  &  $8.97$ & $1.18$ \nl 
\enddata
\tablenotetext{a}{$\sigma^{\rm N}, \sigma^{\rm F}, \sigma^{\rm S}, 
\sigma^{\rm BBKS}$, and 
$\sigma^{\rm EH}$ are same as Table 1.}
\end{deluxetable}

\end{document}